\documentclass[sigconf]{acmart}
\usepackage[ruled,vlined]{algorithm2e}
\usepackage{algpseudocode}
\usepackage{enumitem}
\usepackage{makecell}
\usepackage{multirow}
\usepackage{subfigure}
\usepackage{footnote}
\usepackage{threeparttable}
\usepackage{pifont}
\usepackage{colortbl}
\usepackage{flushend}
\usepackage{tcolorbox}
\tcbuselibrary{breakable,skins}

\usepackage{xcolor}

\colorlet{light-gray}{gray!45!white}
\colorlet{dark-green}{green!70!black}

\usepackage{marginnote}

\algnewcommand{\LineComment}[1]{\State \(\triangleright\) #1}

\usepackage{listings}

\definecolor{codegreen}{rgb}{0,0.6,0}
\definecolor{codegray}{rgb}{0.5,0.5,0.5}
\definecolor{backcolor}{RGB}{245,248,250}
\definecolor{emph}{RGB}{166,88,53}
\definecolor{nightblue}{RGB}{9,49,105}
\definecolor{keywords}{RGB}{207,33,46}
\definecolor{lightpurple}{RGB}{130,81,223}

\lstdefinestyle{mystyle}{
    backgroundcolor=\color{backcolor},
    commentstyle=\color{codegreen},
    keywordstyle=\color{keywords},
    stringstyle=\color{nightblue},
    basicstyle=\ttfamily\footnotesize,
    breakatwhitespace=false,         
    breaklines=true,                 
    captionpos=b,                    
    keepspaces=true,      
    numberstyle=\tiny\color{codegray},
    numbers=left,                    
    numbersep=2pt,                  
    showspaces=false,                
    showstringspaces=false,
    showtabs=false,                  
    tabsize=2,
    linewidth=1\columnwidth,
    frame=tb,
}
\newtheorem{hypothesis}{Hypothesis}  
\newtheorem{example}{Example}  

\newcommand{\set}[1]{\{#1\}}                    
\newcommand{\setof}[2]{\{{#1}\mid{#2}\}}        
\newcommand{\argmin}[1]{\mathop{\arg\min}_{#1}} 
\newcommand{\len}[1]{|#1|}                      

\newcommand{\sql}[1]{\textup{\textsf{#1}}}
\newcommand{\ie}{\emph{i.e.},\xspace}
\newcommand{\eg}{\emph{e.g.},\xspace}

\newcommand{\ours}{\textsf{Nirvana}\xspace}
\AtBeginDocument{%
  }

\setcopyright{acmlicensed}
\copyrightyear{2018}
\acmYear{2018}
\acmDOI{XXXXXXX.XXXXXXX}
\acmConference[Conference acronym 'XX]{Make sure to enter the correct
  conference title from your rights confirmation email}{June 03--05,
  2018}{Woodstock, NY}
\acmISBN{978-1-4503-XXXX-X/2018/06}

\begin{document}

\title[Beyond Relational: Semantic-Aware Multi-Modal Analytics with LLM-Native Query Optimization]{\texorpdfstring{Beyond Relational: Semantic-Aware Multi-Modal Analytics \\with LLM-Native Query Optimization}{Beyond Relational: Semantic-Aware Multi-Modal Analytics with LLM-Native Query Optimization}}

\author{Junhao Zhu}
\affiliation{%
  \institution{Zhejiang University}
  \city{Hangzhou}
  \country{China}
}
\email{zhujunhao@zju.edu.cn}

\author{Lu Chen}
\affiliation{%
  \institution{Zhejiang University}
  \city{Hangzhou}
  \country{China}
}
\email{luchen@zju.edu.cn}

\author{Xiangyu Ke, Ziquan Fang}
\affiliation{%
  \institution{Zhejiang University}
  \city{Hangzhou}
  \country{China}
}
\email{{xiangyu.ke, zqfang}@zju.edu.cn}


\author{Tianyi Li}
\affiliation{%
  \institution{Aalborg University}
  \city{Aalborg}
  \country{Denmark}
}
\email{tianyi@cs.aau.dk}

\author{Yunjun Gao}
\affiliation{%
  \institution{Zhejiang University}
  \city{Hangzhou}
  \country{China}
}
\email{gaoyj@zju.edu.cn}

\author{Christian S. Jensen}
\affiliation{%
  \institution{Aalborg University}
  \city{Aalborg}
  \country{Denmark}
}
\email{csj@cs.aau.dk}

\renewcommand{\shortauthors}{Zhu et al.}

\begin{abstract}

    Multi-modal analytical processing has the potential to transform applications in e-commerce, healthcare, entertainment, and beyond. However, real-world adoption remains elusive due to the limited ability of traditional relational query operators to capture query semantics. 
    The emergence of foundation models, particularly the large language models (LLMs), opens up new opportunities to develop flexible, semantic-aware data analytics systems that transcend the relational paradigm.

    We present \ours, a multi-modal data analytics framework that incorporates programmable semantic operators while leveraging both logical and physical query optimization strategies, tailored for LLM-driven semantic query processing. 
    \ours addresses two key challenges. First, it features an agentic logical optimizer that uses natural language-specified transformation rules and random-walk-based search to explore vast spaces of semantically equivalent query plans --- far beyond the capabilities of conventional optimizers. 
    Second, it introduces a cost-aware physical optimizer that selects the most effective LLM backend for each operator using a novel improvement-score metric.
    To further enhance efficiency, \ours incorporates computation reuse and evaluation pushdown techniques guided by model capability hypotheses. 
    Experimental evaluations on three real-world benchmarks demonstrate that \ours is able to reduce end-to-end runtime by 10\%–85\% and reduces system processing costs by 76\% on average, outperforming state-of-the-art systems at both efficiency and scalability.
\end{abstract}

\maketitle
\section{Introduction}\label{sec:introduction}
Semantics-aware multi-modal data analytics has long been an attractive vision, with the potential to power important applications across domains such as e-commerce, healthcare, and entertainment. 
However, despite its promise, reliable deployment in production environments is still lacking. One major impediment is the rigidity of existing analytical processing systems, whose semantics-unaware operators often struggle to support the querying of unstructured and heterogeneous data.
Fortunately, the emergence of foundation models offers new opportunities, with Large Language Models (LLMs) demonstrating remarkable general-purpose reasoning capabilities, such as document understanding and question answering. 
This opens up new possibilities: the realization of flexible, semantic-aware multi-modal data analytics systems that hold the potential to fundamentally transform analytics by leveraging the complementary strengths of foundation models and large-scale data systems.
On the one hand, LLMs bring semantic understanding across modalities --- including text, images, and audio --- into the core of data analytics. On the other hand, system engineering enables these models to be deployed with greater efficiency and transparency.

As a type of compound AI system, multi-modal semantic analytical processing systems can transcend monolithic models by being envisioned to offer the following appealing features:

\noindent(1) \textbf{High-Quality \& Versatile Data Analytics.} 
These systems are designed to support a wide range of analytics tasks, including descriptive, predictive, and prescriptive analytics, while delivering high-quality results that are robust and production-ready. 

\noindent(2) \textbf{Efficiency \& Scalability.} 
Like relational database systems, semantics-aware multi-modal 
analytics systems must be capable of processing large-scale, data-intensive workloads efficiently and cost-effectively, scaling to tens of millions of records and beyond. 

\noindent(3) \textbf{Automatic Optimization.} 
Inspired by traditional relational query optimization, these systems should provide transparent, end-to-end optimization of semantic analytical workflows. Users should be able to specify {\em what} they want to achieve, without worrying about {\em how} to achieve it.

\begin{figure*}[t]
    \centering
    \includegraphics[width=\textwidth]{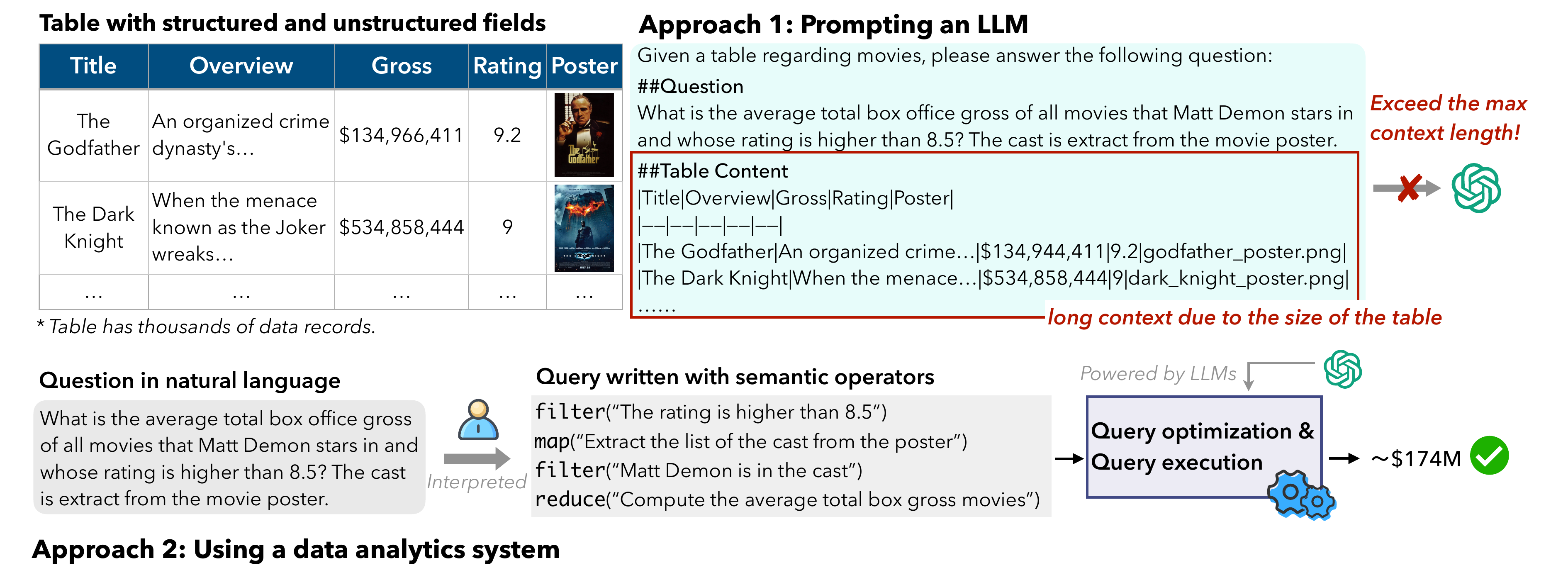}
    \caption{Example of performing data analytics on a multi-modal movie table by (1) directly prompting an LLM or (2) using a data analytical processing system.}
    \label{fig:example}
    \vspace{-5mm}
\end{figure*}

\begin{example}[Prompt vs. Compound System]
    Figure~\ref{fig:example} illustrates two approaches to answering the query, ``Compute the average total box office gross of all movies that Matt Damon stars in and whose IMDb rating is higher than 8.5,'' over a multi-modal table, where the cast must be extracted from movie posters. 
    The first approach (upper part) constructs a single prompt by concatenating the query with the full table contents and submits it directly to an LLM. 
    In realistic scenarios --- with tables containing hundreds or thousands of rows --- this method fails because the prompt’s token length exceeds the model’s maximum context window, causing truncation or outright rejection.
    The alternative approach leverages a semantic analytical processing system that exposes a set of programmable operators powered by LLMs. The system decomposes the original query into a sequence of atomic semantic tasks --- for example, extracting the cast from posters, filtering out movies with IMDb ratings below 8.5 or without Matt Damon in the cast, etc. Each task is executed on appropriately sized batches of records, ensuring that no single LLM invocation exceeds the context limit. Additionally, the system employs various system-level optimizations, such as parallel execution, to make the analytics more efficient and cost-effective.
\end{example}

The recent surge of interest in semantic analytical processing has given rise to two complementary research strands. 
First, several efforts integrate AI-powered operators into existing SQL engines. For example, ThalamusDB~\cite{jo2024thalamusdb} introduces semantic filtering backed by pretrained models into DuckDB. CAESURA~\cite{urban2024caesura} augments traditional SQL operators (\eg joins and aggregations) with semantic ones (\eg TextQA and VisualQA) for question answering over text and images. 
While these extensions unlock new functionality, they exhibit low compatibility with the query optimizers in existing systems. 
Second, a growing body of studies target ground-up, purpose‑built semantic analytical systems~\cite{liu2024palimpzest, patel2024lotus, shankar2024docetl, anderson2024aryn, wang2025aop}. 
These systems expose declarative, natural‑language‑parameterized transformations --- such as the `filter' and `map' operators used in Figure~\ref{fig:example} --- alongside traditional relational operators. Notably, they incorporate system‑level optimizations: for example, Palimpzest~\cite{liu2024palimpzest} adapts the Cascades optimization framework to generate semantic query execution plans that strike a Pareto-optimal balance among quality, latency, and cost. 
Despite this progress, fundamental questions remain unanswered. Classical cost estimation falls short when the selectivity and performance of operators depend on LLM inference behavior. 
Moreover, {\em analysis quality} emerges as a first‑class optimization objective due to the probabilistic nature and occasional hallucinations of foundation models. 
\emph{Beyond simply extending SQL or reusing relational optimizers, what new optimization paradigms can unlock the full promise of multi‑modal semantic processing?}

We present \ours, a novel multi-modal analytical processing framework that offers programmable semantic operator interfaces and introduces new strategies for system-level optimization. Rather than sidelining existing efforts on semantic processing, the goal of \ours is to present complementary optimization opportunities for semantic processing that existing systems overlook. In addition to implementing semantic operators that enable versatile data analytics, \ours contributes two new optimization designs, driven by the following key questions:

\noindent\textbf{\textit{Question 1: What makes semantic query optimization different from traditional query optimization?}} Unlike traditional relational operators, semantic operators invoke LLM inference to process multi-modal data, which is expensive and dominates the processing cost due to the high computational cost associated with LLM inference. Therefore, the prime object of query optimization is to reduce the number of LLM calls. Borrowing from relational DBMSs, \ours adopts transformation-based logical plan optimization (semantic query rewriting), employing traditional rules like predicate pushdown and new rules tailored to semantic analytics, \eg operator fusion and non-LLM replacement. These transformations require the optimizer to understand query semantics that existing query optimizers cannot handle. LLMs are a silver bullet for semantic query optimization, as they have been shown to be capable of solving complex tasks such as workflow orchestration~\cite{shen2023hugginggpt, zhuge2024gptswarm, zhang2025aflow, hu2024adas}. \ours employs an agentic logical plan optimizer that leverages LLMs to explore semantically equivalent but more efficient plans using a random walk-based search framework. However, we note that, in relational databases, query optimization overhead is negligible compared to query execution (\eg milliseconds for terabytes of data according to the TPC-H benchmark~\cite{tpch}), while LLM calls can take seconds to complete. Given the high cost of LLMs, \textbf{\textit{is it worth using LLMs for query optimization?}} Our answer is yes. Unlike the query optimizations in relational databases that is based on data statics,  \textit{semantic query optimization can benefit markedly from the the capabilities of LLM, despite the associated costs}. Our logical plan optimizer can yield larger savings (in both runtime and monetary cost) compared to the optimization overhead. Our empirical studies show that the logical plan optimizer can reduce end-to-end execution time by $40.6\%$ and processing price by $25.0\%$ on average, respectively.

\noindent\textbf{\textit{Question 2: Given logical plans for semantic data analytics, do physical plan optimizations exist?}} Service providers usually offer a variety of LLM options. For instance, OpenAI offers three sizes of its GPT-4.1 series: GPT-4.1, GPT-4.1 mini, and GPT-4.1 nano. These models exhibit heterogeneous performance and costs, where a more powerful LLM is inevitably more expensive. Each semantic operator in a plan can choose a variety of models to use. Among all LLM options, some are overkill for simple analytics requests, while others may underperform on complex ones. \ours defines a physical plan as a choice of execution strategy (\ie models in this work) for operators in a logical plan, and we apply physical plan optimization through model selection. Jo et al.~\cite{jo2024smart} proposed Smart, a model selection method, which leverages a mix of LLMs to provide similar levels of output quality as the most powerful LLM but at a much lower expense. However, Smart only supports selecting LLMs for a single operator in a fixed task. \ours supports general query plans, which optimizes physical plans by assigning the most cost-effective model to each operator based on an improvement score, a novel measure of model cost-effectiveness. To improve optimization efficiency, we introduce two techniques, \emph{computation reuse} and \emph{evaluation pushdown}, to eliminate unnecessary computations. The empirical study (Section~\ref{sec:experiment}) finds that these optimizations reduce optimization overhead by $4\times$ compared to Smart, while maintaining system effectiveness.

In summary, the key contributions are as follows:
\begin{itemize}[topsep=0pt]
    \item We discuss differences between semantic data analytics and relational data analytics, motivating an agent-based logical plan optimizer that uses a random walk search framework and an LLM-as-a-judge plan evaluator for logical plan exploration and exploitation.
    \item We design a physical plan optimizer that selects the most cost-effective LLM backend per operator in a logical plan, minimizing inference costs while preserving result quality.
    \item The core of the physical plan optimizer is a model selection algorithm based on the novel concept of an improvement score. To reduce optimization overhead, we propose optimization techniques --- computation reuse and evaluation pushdown --- with a model capability hypothesis.
    \item Extensive experiments on three real-world multi-modal benchmarks show that \ours can reduce end-to-end runtime by $10\%$–-$85\%$ and monetary cost by $76\%$, compared to state-of-the-art systems. Each component plays its unique role in facilitating system's efficiency and scalability.
\end{itemize}

The remainder of the paper is organized as follows. Section~\ref{sec:overview} provides an overview of \ours. Section~\ref{sec:logical_plan} presents the logical plan optimizer in \ours. Section~\ref{sec:physical_plan} presents the physical plan optimizer. We report experimental results in Section~\ref{sec:experiment}. We discuss related work in Section~\ref{sec:related_work} before concluding and discussing future directions in Section~\ref{sec:conclusion}.
\ours is open-sourced at \textcolor{blue}{\url{https://github.com/JunHao-Zhu/nirvana}}.
\begin{table*}[t]
    \centering
    \setlength{\abovecaptionskip}{1mm}
    \caption{Semantic operators currently supported in \ours.}
    \label{tab:operator}
    \renewcommand{\arraystretch}{1.1}
    \begin{tabular}{p{1.5cm}|p{7cm}|p{8cm}}
    \toprule[1pt]
    \textbf{Operator} & \textbf{Definition} & \textbf{Usage Example} \\ \hline
    \sql{map} & Applies an LLM to perform a transformation on a column, producing a new column as output & \sql{df.semantic\_map(user\_instruction="NL-specified transformation", input\_column="colA", output\_column="colB")} \\ \hline
    \sql{filter} & Uses an LLM to evaluate a natural language predicate on a column & \sql{df.semantic\_filter(user\_instruction="NL-specified predicate", input\_column="colA")} \\ \hline
    \sql{reduce} & Aggregates values in a column based on an NL-specified reduction function & \sql{df.semantic\_reduce(user\_instruction="NL-specified reducer", input\_column="colA")} \\ \hline
    \sql{rank} & Ranks values in a column according to an NL-specified ranking function & \sql{df.semantic\_rank(user\_instruction="NL-specified ranker", input\_column="colA")} \\
    \bottomrule[1pt]
    \end{tabular}
\end{table*}
\vspace{-2mm}

\section{System Overview}\label{sec:overview}
We first provide background on the data model and analytical queries supported by \ours. Then, we define the notions of logical and physical plans in the context of \ours and outline the optimizations involved.

\subsection{Background: Data Model \& Operators}
\ours supports a data model consisting of tables with both structured and unstructured fields. Unstructured fields may contain heterogeneous content such as text, documents, images, or audio. Figure~\ref{fig:example} presents an example of a movie dataset containing numeric, textual, and image fields. Internally, \ours represents unstructured fields as text that store file paths pointing to remote locations (\eg web URLs or object storage such as S3~\cite{s3}) where the unstructured data resides.

To support flexible and versatile semantic data analytics, \ours offers a suite of semantic operators. Table~\ref{tab:operator} summarizes the operators currently implemented in \ours: The \sql{map} operator applies an LLM-driven transformation to a target column, producing a new derived column. The \sql{filter} operator retains records based on a natural language predicate. The \sql{reduce} operator applies a many-to-one aggregation across multiple data in a target column using a user-specified reducer expressed in natural language (NL). The \sql{rank} operator assigns rankings to input values using an NL-defined ranking criterion. While Table~\ref{tab:operator} lists the currently supported operators, \ours is extensible with support for additional operators such as \sql{join} and \sql{group by} in this work. Users can construct their own analytical queries tailored to their data and tasks by employing semantic operators in \ours. We describe the system as multi-modal because it operates over multi-modal relational data, and semantic operators can interpret queries and generate results involving different modalities. The following is an example query to analyze common characteristics of crime movies with ratings between 8.5 and 9:

\begin{lstlisting}[language=Python, label={lst:query_example}, caption=Query example.]
import nirvana as nvn
df = nvn.DataFrame(movie_data)
df.map(user_instruction="According to the movie overview, extract the genre of each movie.", input_column="Overview", output_column="Genre")
df.filter(user_instruction="The rating is higher than 7.", input_column="Rating")
df.filter(user_instruction="The rating is lower than 8.", input_column="Rating")
df.filter(user_instruction="The movie belongs to crime movies.", input_column="Genre")
df.reduce(user_instruction="Summarize the common points of their stories.", input_column="Overview")
\end{lstlisting}
\vspace{-3mm}

\begin{figure*}[t]
    \centering
    \includegraphics[width=\textwidth]{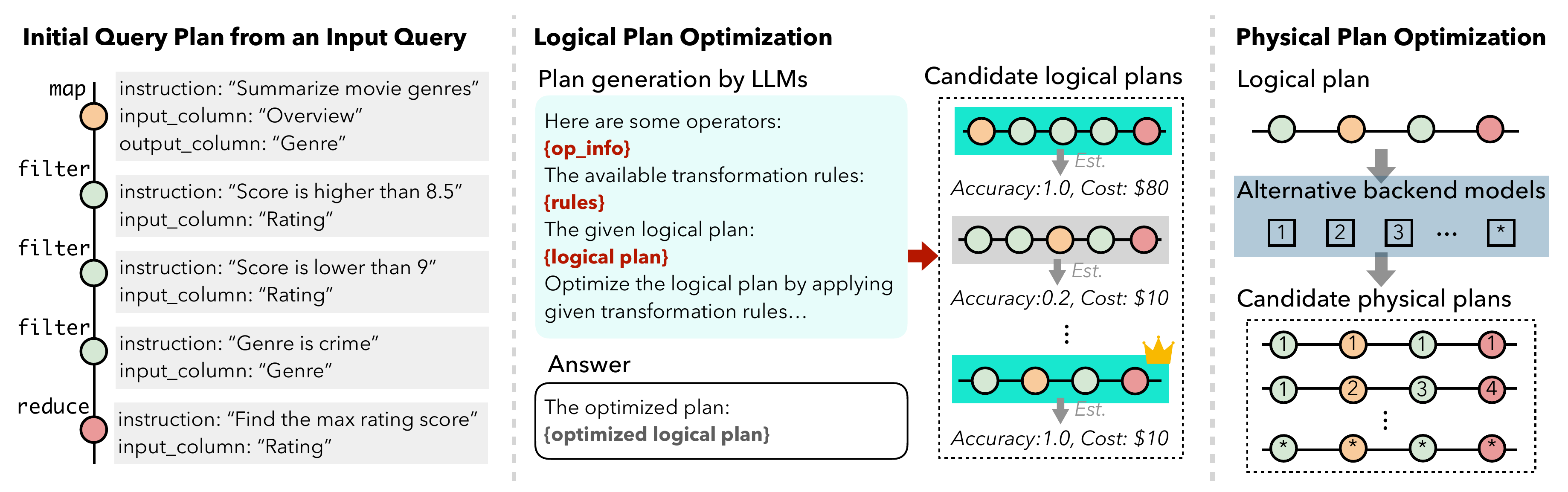}
    \vspace{-6mm}
    \caption{System overview of \ours. Given an initial query plan, the logical optimizer samples and rewrites the logical plan to identify a (near-)optimal plan in the logical plan optimization phase. Then, the physical optimizer assigns the most cost-effective model to each operator in the logical plan.}
    \vspace{-3mm}
    \label{fig:overview}
\end{figure*}

\subsection{Query Optimization}
At present, optimization in \ours aims to minimize system cost and latency under an accuracy constraint. 
This reflects common approximate-query practice where small, bounded deviations from the oracle result are acceptable in exchange for cost/latency improvements. In \ours, a user-specified query is compiled into data lineage (a directed acyclic graph), which serves as a logical plan in this paper.
A logical plan captures the execution order of semantic operators, along with their predicates or instructions and their scopes of application, as illustrated on the left side of Figure~\ref{fig:overview}, where a logical plan is derived from the query in Listing~\ref{lst:query_example}. A corresponding physical plan specifies the actual execution engines (e.g., computing functions or AI models) used to implement each operator in the logical plan. Next, \ours offers logical and physical plan optimizations. Inspired by the Spark SQL system~\cite{armbrust2015sparksql}, \ours separates query optimization into logical plan optimization and physical plan optimization.

While logical plan optimization in \ours mirrors and extends traditional rule-based approaches, \ours devises a key innovation --- an agentic optimizer --- and introduce several transformation rules specific to semantic processing (\ie operator re-ordering, operator fusion, and operator replacement). As indicated in Figure~\ref{fig:overview}, the agentic optimizer explores multiple candidate logical plans and estimates their result quality (measured by accuracy) and execution cost. It then selects the plan that minimizes the cost while meeting a quality constraint. The details of the logical plan optimizer are covered in Section~\ref{sec:logical_plan}. As shown to the right part in Figure~\ref{fig:overview}, following the logical plan optimization, in \ours, the physical plan optimizer assigns appropriate backend engines (\ie LLMs) to each operator in the selected logical plan. The goal is to choose the most cost-effective configuration under the quality constraint. Details are presented in Section~\ref{sec:physical_plan}. Note that, the optimizers are modality-agnostic: optimizations apply uniformly across data types.

\section{Logical Plan Optimizer}\label{sec:logical_plan}
We begin by presenting the logical plan optimizer, which applies transformation rules to the initial (user-specified) plan to produce an optimized logical plan.

\begin{algorithm}[t]\small
\caption{Logical plan optimization}\label{algo:logical_optim}
\LinesNumbered
\DontPrintSemicolon
    \KwIn{Data samples $D_{s}$, an initial plan $p_0$, an plan rewriter $m$, the number of iterations $N_{\mathit{max}}$, error tolerance $\epsilon$}
    \KwOut{the optimized logical plan $p^{*}$}
    Initialize the set of candidate plans $\mathcal{P}\leftarrow\set{p_0}$\;
    Initialize the best logical plan $p^{*}\leftarrow p_0$\;
    \For{$i\leftarrow 1$ to $N_{\mathit{max}}$} {
    $p_i\leftarrow\mathit{sample}(\mathcal{P})$\textcolor{light-gray}{\Comment{Sample a plan from candidates}}\;
    $p^{+}\leftarrow\mathit{rewrite}(p_i, m)$\;
    $p^{+}.\mathit{acc}, p^{+}.\mathit{cost}\leftarrow\mathit{evaluate}(p^{+}, p_{0}, D_s)$\textcolor{light-gray}{\Comment{Estimate accuracy and cost of the optimized plan}}\;
    \If{$p^{*}.\mathit{acc}\ge\epsilon\cap p^{+}.\mathit{cost}\le p_{i}.\mathit{cost}$} {
        $\mathcal{P}\leftarrow \mathcal{P}\cup\set{p^{*}}$;
    }
    }
    $p^{*}\leftarrow\argmin{p}\set{p.\mathit{cost}\mid p\in\mathcal{P}}$\;
\end{algorithm}

\subsection{Optimization Workflow}
\begin{figure}[t]
    \centering
    \includegraphics[width=\linewidth]{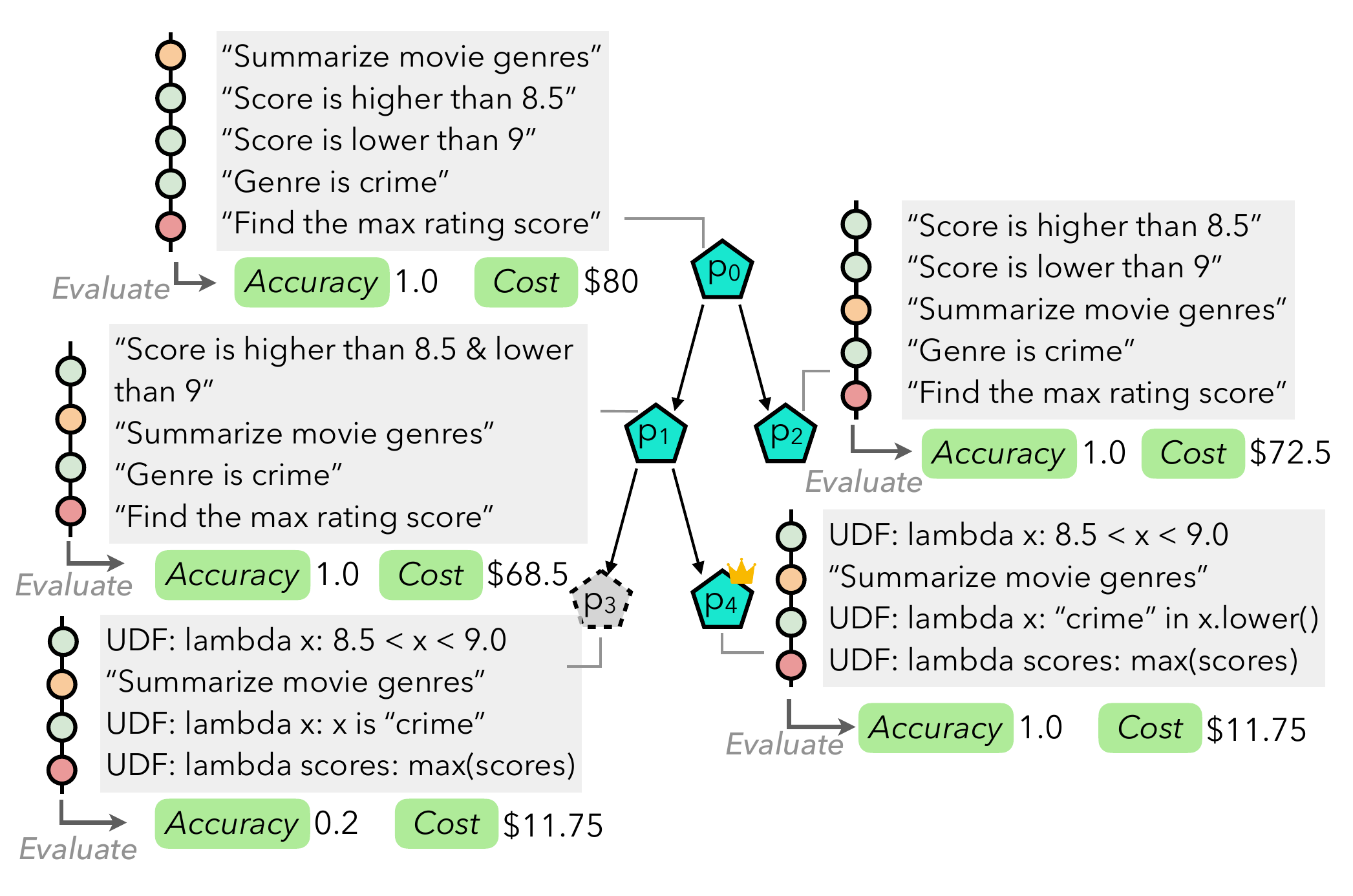}
    \vspace{-4mm}
    \caption{A running example of logical plan optimization where, each node (a logical plan) is sampled by random walk.}
    \label{fig:logical_plan}
    \vspace{-3mm}
\end{figure}

Algorithm~\ref{algo:logical_optim} outlines the skeleton of the logical plan optimizer, and Figure~\ref{fig:logical_plan} illustrates the workflow with a running example. 

\noindent\textbf{Initialization.} Given a user-specified plan as input, the optimizer initializes the set $\mathcal{P}$ of candidate logical plans with an empty set and sets the best logical plan $p^{*}$ to the input plan $p_0$ (lines 1--2). 

\noindent\textbf{Plan rewrite.} In each iteration, a plan is sampled from the candidate set for rewriting (line 4). We adopt a random walk approach~\cite{grover2016node2vec} to sample a plan from the candidate set. Each plan is assigned a probability such that all candidates have a chance to be sampled. The random walk spans a tree, and each node represents a logical plan, as shown in Figure~\ref{fig:logical_plan}, where child nodes are generated from their parent node by rewriting plans. To avoid getting stuck in a local optimum, each plan is sampled from a distribution of a mixture of a uniform distribution and a cost-based weighted probability distribution. Specifically, a plan $p_i$ is sample with the probability:
\begin{equation}\label{eq:sample}
    Pr(p_i)=\lambda\cdot\frac{1}{\len{\mathcal{P}}}+(1-\lambda)\cdot\frac{\exp(\mathit{cost}_{\mathit{max}} - p_i.\mathit{cost})}{\sum_{p_j\in\mathcal{P}}\exp(\mathit{cost}_{\mathit{max}} - p_j.\mathit{cost})},
\end{equation}
where $\lambda$ (set to $0.2$) controls the tendency to either continue optimizing the currently best observed plan or to explore new branches. This random walk search procedure interpolates between breadth-first search and depth-first search, thereby balancing plan exploration and plan exploitation.

The selected plan is then rewritten by an LLM rewriter using transformation rules expressed in natural language (line 5). The concept of LLM-driven plan rewriting has been validated and adopted in various domains, including in workflow generation~\cite{zhuge2024gptswarm, zhang2025aflow} and document analytics~\cite{shankar2024docetl}.

\noindent\textbf{Plan evaluation.} Once the sampled plan is rewritten, we check for semantic equivalence and evaluate the plan's cost. If the new plan is semantically equivalent to the input plan and incurs lower cost, it is added to the candidate set (lines 6--8).

Verifying semantic equivalence is quite tricky, especially for semantic queries. There are two representative query verification approaches: (1) \textit{Formal verification}: converts SQL queries into logic formulas and checks the equivalence via algebraic solutions, \eg SQLSolver~\cite{ding2023sqlsolver}. 
(2) \textit{Execution consistency}: check if the execution result of a query matches that of the corresponding rewritten query, \eg SQLDriller~\cite{yang2025sqldriller}.
Formal equivalence checking is attractive but difficult to apply in semantic analytics because many natural-language-driven operators lack straightforward logical encodings. Although one can use LLMs to perform formal equivalence checking, it creates a circular trust issue that it uses the same LLM to both rewrite and verify.
To ensure an \emph{objective} verification, \ours performs execution consistency checking that the execution results of rewritten plans are compared with those of the original plan.
However, result equivalence can also be vague in semantic analytics, as LLM-executed semantic operators may produce syntactically different but semantically identical outputs for the same inputs. To address this, we use LLM-as-a-judge~\cite{zheng2023llmasajudge} as an equivalence verifier. We execute both the input plan and the rewritten plan on a data sample. With the outputs from both logical plans, we prompt an LLM to rate the similarity between the two plan outputs on a scale from 0 (completely different) to 10 (exactly same) based on the their semantic consistency. The normalized rating score is used as the plan's accuracy.

For plan cost, the dominant factor, the overall running time and the monetary cost for LLM inference are typically proportional to the number of LLM calls (\ie the number of data points to process) and the prompt length per data processing. Our cost estimator tracks the number of processed data items and prompt lengths per operator. The total cost is computed as the sum of all operator costs. We assign a selectivity parameter to each operator to monitor changes in the number of data points to be processed throughout the plan. By default, \sql{filter} has a selectivity of 0.5, \sql{reduce} has a selectivity of 0, and the other operators have one of 1. We track changes to operators in the plan and adjust their selectivity accordingly. For instance, when multiple operators (\eg two or three filters) are merged, the new operator's selectivity is adjusted accordingly (\eg reduced to half or one third, depending on how many filter operators are merged).

Although our system uses constant selectivity assumptions by default for now, operator selectivity can also be estimated using cardinality estimation techniques, similar to those in relational DBMSs.
However, semantic data analytics depends not only on data distribution but also on natural language instructions, motivating a new line of research on semantic cardinality estimation.
\ours can seamlessly integrate such semantic estimators once they become available. For instance, consider two \sql{filter} operators in a query plan. Under the default optimizer, their order cannot be determined as they have identical selectivities. With a semantic cardinality estimator, however, the optimizer can prioritize execution by pushing down the \sql{filter} with the higher selectivity, thereby reducing overall query cost. In addition, it also works for determining a join order.

At the end of the optimization, the logical plan optimizer returns the lowest-cost plan that satisfies the accuracy constraint (line 9). In the case of Figure~\ref{fig:logical_plan}, both $p_3$ and $p_4$ have low costs, but $p_3$ is discarded because it fails the semantic equivalence check.

\subsection{Transformation Rules}\label{sec:rules}
Traditional query optimizers based on frameworks like Volcano~\cite{graefe1993volcano} and Cascades~\cite{gradfe1995cascades} face several limitations in the context of semantic data analytics:
\begin{enumerate}[label=(\arabic*), topsep=0pt, leftmargin=*]
    \item \emph{Missing optimization opportunities.} Traditional rule-based systems provide deterministic, hard-coded transformations that are hard to extend at query-time and hard to offer flexible, query-specific semantic interpretations and rewrites that are difficult to pre-enumerate.
    \item \emph{Limited extensibility.} Transformation rules in existing query optimizers are usually hard-coded and tightly coupled with system internals~\cite{wang2024lapuda}. Integrating a new transformation rule into a commercial or open-source database system is a non-trivial undertaking.
\end{enumerate}

Our agentic logical plan optimizer addresses these issues. First, it leverages the semantic understanding of LLMs to enable more aggressive and intelligent plan rewrites. Second, the transformation rules are expressed in natural language, decoupling them from system internals.

The following three transformation rules are applied in our logical optimizer. Additional rules, formulated in natural language, can be added easily.
\begin{itemize}[topsep=0pt,leftmargin=*]
    \item \textbf{Filter pushdown.} This transformation, widely used in existing DBMSs, moves a filter operator that does not rely on results of preceding operators to an earlier stage in the plan.
    \item \textbf{Operator fusion.} This transformation merges multiple operators on the same field into one operator. To maintain semantic equivalence, the predicate (or user instruction) will be rewritten for the new operator. For example, in Figure~\ref{fig:logical_plan}, two filter operations with predicates ``Score is higher than 8.5'' and ``Score is lower than 9'' can be merged into a single filter with predicate ``Score is higher than 8.5 and lower than 9'', yielding fewer LLM calls in processing a query.
    \item \textbf{Non-LLM replacement.} This transformation replaces an operator's NL instruction with an equivalent compute function. For example, in Figure~\ref{fig:logical_plan}, ``Score is higher than 8.5 and lower than 9'' can be interpreted as a comparison function: \sql{8.5 < score < 9}.
\end{itemize}

\begin{algorithm}[t]\small
\caption{Naive backend model selection}\label{algo:physical_optim}
\LinesNumbered
\DontPrintSemicolon
    \KwIn{Data samples $D_{s}$, an operator $o$, a model set $\mathcal{M}=\set{m_1,m_2,m_3,m^{*}}$, an improvement margin $\Delta I_{\mathit{min}}$}
    \KwOut{the most cost-effective model $m_o$}
    Initialize improvement scores $I_{\mathit{last}}\leftarrow 0, I_{\mathit{curr}}\leftarrow 0$\;
    Initialize the most cost-effective model $m_o\leftarrow m_1$\;
    $R_{m_1}, R_{m^{*}}\leftarrow \setof{m_1(x)}{x\in D_s}, \setof{m^{*}(x)}{x\in D_s}$\textcolor{light-gray}{\Comment{Results of running $m_1$ and $m^{*}$ on $D_s$ used for computing the model cost-effectiveness estimation}}\;
    \For{$m$ in $\mathcal{M}$} {
    \If{$m = m_1$} {
        $I_\mathit{curr}\leftarrow 0$\;
        \textbf{continue}\;
    }
    $R_{m}\leftarrow \setof{m(x)}{x\in D_s}$\;
    $I_{\mathit{curr}}\leftarrow$ estimate model cost-effectiveness by computing the improvement score $I_{m_1\rightarrow m}(D_s,m^{*})$ on $R_{m_1}, R_{m}$, and $R_{m^{*}}$\;
    \If{$I_{\mathit{curr}} - I_{\mathit{last}} \ge \Delta I_{\mathit{min}}$} {
        $m_o\leftarrow m$\;
        $I_{\mathit{last}}\leftarrow I_{\mathit{curr}}$\;
    }
    }
\end{algorithm}

\subsection{Local Rewrite Model}
The cost overhead is one of the biggest obstacles when using LLMs for plan optimization, as each LLM call is expensive in both latency and price. Especially when using LLM services provided by cloud vendors, most of the time is wasted on network communication. One means of mitigating this problem is to use a local rewrite model as a substitute.

\noindent\textbf{Training data collection.} A dataset to be used for training a local rewrite model consists of pairs of un-optimized and rewritten logical plans. To assemble such pairs, we first use an LLM to convert natural language data analytics questions into analytical queries using the semantic operators in \ours, providing the question and operator interface documentation. Then, we use the LLM with transformation rules to rewrite the logical plan compiled from the generated queries in a last step. Collecting the training set, we fine-tune a small LLM (e.g., Qwen-3-8B, LLaMA-3-8B) using LoRA~\cite{hu2022lora} in an offline setting to obtain the local rewriter.

\noindent\textbf{Model inference.} In the inference phase, the sampled logical plan is fed to the local rewrite model to generate an optimized version, \ie replacing the cloud LLM service with the local model in the optimization workflow. As cloud vendors improve LLM services in latency and others features, the local model's inference needs to be carefully implemented to achieve the same effect. For instance, to further accelerate generation, LLM serving systems (\eg vLLM~\cite{kwon2023vllm}, SGLang~\cite{zheng2024sglang}) are recommended for local deployment.
\section{Physical Plan Optimizer}\label{sec:physical_plan}
In this work, a physical plan captures the assignment to semantic operators of backend models used as execution engines. Different models with varying scales offer different levels of intelligence, speed, and cost. Selecting the most appropriate model for each operator is essential for cost-effective data analytics, which is the goal of physical plan optimization. For readability, we discuss a case with a search space of four models\footnote{Why do we choose a four-model setting to discuss? It can cover most use practices, where users may choose among a Tall-sized model (\eg BERT), a Gronde-sized model (\eg LLaMA3-8B), a Venti-sized model (\eg LLaMA3-70B), and a Trenta-sized model (\eg GPT-4.1). The discussion generalizes to cases with more models.}, \ie $\mathcal{M}=\set{m_1, m_2, m_3, m^{*}}$. These models have the following properties:
\begin{equation*}
\begin{split}
    \text{Intelligence (from weakest to strongest):}\quad m_1\prec m_2\prec m_3\prec m^{*} \\
    \text{Cost (from cheapest to most expensive):}\quad m_1\succ m_2\succ m_3\succ m^{*}
\end{split}
\end{equation*}
Model $m_1$ is the weakest but fastest (and cheapest), while $m^{*}$ is the most powerful but slowest (and most expensive).

\begin{figure}[t]
    \centering
    \includegraphics[width=\linewidth]{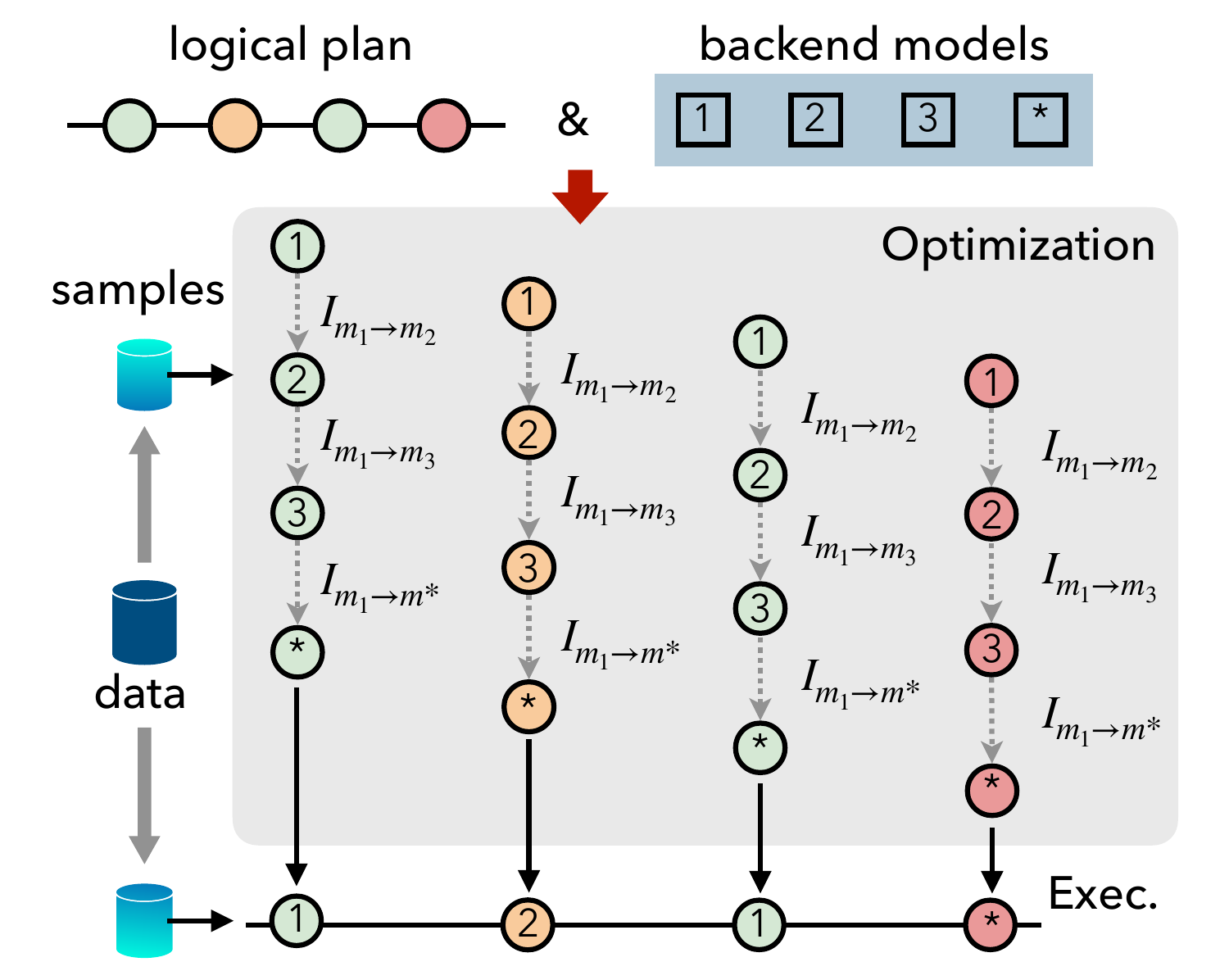}
    \vspace{-3mm}
    \caption{Overview of physical plan optimization.}
    \label{fig:physical_plan}
    \vspace{-3mm}
\end{figure}

\subsection{Optimization Workflow}
Given a logical plan with $N$ operators and a choice of $M$ models per operator, the search space of feasible physical plans is of size $N^M$, yielding a search space that is too large to efficiently find an optimal physical plan. Hence, we constrain the search space by making the following assumption:

\begin{hypothesis}[Operator Independence]\label{hypo:op_independence}
    Suppose that a model $m_a$ is more intelligent and costly than a model $m_b$. Then, choosing $m_a$ over $m_b$ as the execution engine for an operator in a plan improves the overall result quality (but at higher cost), regardless of the models allocated to other operators in the plan.
\end{hypothesis}

The hypothesis overlooks potential error propagation within the workflow --- errors produced by one operator may cascade and amplify through subsequent operators. Fortunately, since inter-op interactions are handled by the logical optimizer while the physical optimizer considers only intra-op optimization, the impact of hypothesis~\ref{hypo:op_independence} is minimal. We adopt a greedy solution for model selection.
Figure~\ref{fig:physical_plan} provides an overview of the physical plan optimization process. The system selects the most cost-effective model for each operator from a pool of candidates. In the subsequent execution phase, the chosen model is used to process the remaining data.
Algorithm~\ref{algo:physical_optim} captures the model selection procedure for a single operator, presenting a naive baseline that forms the foundation for our optimizer. Given an operator in the logical plan, we evaluate the cost-effectiveness of each candidate model by computing an improvement score $I_{m_1 \rightarrow m}$ over a sampled dataset $D_s$. This score quantifies the gain in output quality obtained by replacing the baseline model $m_1$ with a candidate model $m$ as the execution backend. If the score exceeds a user-defined threshold $\Delta I_{\mathit{min}}$, the model is deemed sufficiently beneficial and is assigned to the operator. To estimate model cost-effectiveness, one always needs to run models $m_1$, $m$, and $m^{*}$ on the entire sampled dataset, which can be expensive. To reduce this cost, we introduce several optimization techniques that decrease the amount of data that must be processed.

In the following, we define the improvement score and present acceleration strategies based on it.

\subsection{Improvement Score Computation}
Since the ground-truth result of a query is unknown, we treat the output\footnote{we use terms ``model output'' and ``response'' interchangeably in this paper.} of the most powerful model $m^{*}$ as a proxy for the ground truth. We define the improvement score $I_{m_1\rightarrow m}$ as the performance improvement gained from switching from the backend model $m_1$ to a stronger model $m$. The definition is as follows.
\begin{equation}\label{eq:impr_def}
\begin{split}
    &I_{m_1\rightarrow m_2}(D, m^*)\triangleq \mathbb{E}_{x\sim D}[m_2(x)=m^{*}(x), m_1(x)\neq m_2(x)]\\
    &I_{m_1\rightarrow m_3}(D, m^*)\triangleq \mathbb{E}_{x\sim D}[m_3(x)=m^{*}(x), m_1(x)\neq m_3(x)]\\
    &I_{m_1\rightarrow m^{*}}(D, m^*)\triangleq \mathbb{E}_{x\sim D}[m_1(x)\neq m^{*}(x)],
\end{split}
\end{equation}
where $m(x)$ denotes the response of the model $m$ to an input $x$. And, $m_1(x) =m_2(x)$ (resp. $m_1(x)\neq m_2(x)$) denotes that the outputs from the models $m_1$ and $m_2$ are semantically equal (resp. different). For binary-output operators like \sql{filter}, which are required to return \sql{True} or \sql{False}, the model outputs can be compared directly. For other ad-hoc queries, such as \sql{map} that returns arbitrary values, the model outputs are compared using semantic similarity via semantic embeddings generated by a pre-trained language model, like Sentence-BERT~\cite{reimers2019sbert}.

As shown in Eq.~\ref{eq:impr_def}, an improvement score $I_{m_1\rightarrow m}$ behaves as an expected proportion of inputs from a dataset $D$ to which the responses of model $m$ are consistent with the surrogate ground truth $m^{*}(x)$ while responses of the baseline model $m_1$ are not.

For simplicity, we omit $D$ and $x$, since models always process the same data with respect to a given operator.

Next, to compute exact improvement scores, it is necessary to invoke the most expensive model $m^{*}$ on all inputs. To reduce this potentially large overhead, we make the following transformation that reduces the number of $m^{*}$ invocations.
\begin{equation}\label{eq:impr_compute12}
\begin{split}
    I_{m_1\rightarrow m_2}&\triangleq \mathbb{E}[m_2=m^{*}, m_1\neq m_2]\\
    &=Pr(m_2=m^{*}, m_1\neq m_2)\\
    &=Pr(m_2=m^{*}\mid m_1\neq m_2)Pr(m_1\neq m_2)
\end{split}
\end{equation}
This transformation allows us to execute expensive evaluations $m_2=m^{*}$ on only the subset of records that meet the condition $m_1\neq m_2$, reducing the number of evaluations involving the expensive model, $m^{*}$. We refer to this optimization as ``\emph{evaluation pushdown}'' because its core idea is similar to ``predicate pushdown'' in DBs.

The improvement score $I_{m_1\rightarrow m_3}$ can be transformed in a similar way, \ie $I_{m_1\rightarrow m_3}=Pr(m_3=m^{*}\mid m_1\neq m_3)Pr(m_1\neq m_3)$. However, we can further reduce the number of invocations of the best model $m^{*}$ by applying the law of total probability:
\begin{equation}\label{eq:impr_compute13}
\begin{split}
    &I_{m_1\rightarrow m_3}=Pr(m_3=m^{*}, m_1\neq m_3)\\
    &=Pr(m_3=m^{*}, m_1\neq m_3, m_2=m_3)+Pr(m_3=m^{*}, m_1\neq m_3, m_2\neq m_3) \\
    &=I_{m_1\rightarrow m_2}+Pr(m_3=m^{*}, m_2\neq m_3, m_1=m_2)\\
\end{split}
\end{equation}
The last term in Eq.~\ref{eq:impr_compute13} can be expanded as follows.
\begin{equation*}
\begin{split}
    &Pr(m_3=m^{*}, m_2\neq m_3, m_1=m_2)=\\
    &Pr(m_3=m^{*}\mid m_2\neq m_3, m_1=m_2)Pr(m_2\neq m_3\mid m_1=m_2)Pr(m_1=m_2)
\end{split}
\end{equation*}

From Eq.~\ref{eq:impr_compute13}, the number of $m^{*}$ invocations in the evaluation of $m_3=m^{*}$ can be reduced by the conditions $m_2\neq m_3$ and $m_1=m_2$, where $m_1=m_2$ was already evaluated when computing $I_{m_1\rightarrow m_2}$. We also note that the computing $I_{m_1\rightarrow m_3}$ directly reuses the result of $I_{m_1\rightarrow m_2}$. Therefore, we call this technique ``\emph{computation reuse}''.

These conclusions also apply to the computation of $I_{m_1\rightarrow m^{*}}$. For brevity, we only show the expansion of $I_{m_1\rightarrow m^{*}}$ using the law of total probability:
\begin{equation}\label{eq:impr_compute14}
\begin{split}
    &I_{m_1\rightarrow m^{*}}=Pr(m_1\neq m^{*})\\
    &=Pr(m_1\neq m^{*},m_1=m_2,m_2=m_3)+Pr(m_1\neq m^{*}, m_1=m_2,m_2\neq m_3)\\
    &+Pr(m_1\neq m^{*}, m_1\neq m_2, m_2= m_3)+Pr(m_1\neq m^{*}, m_1\neq m_2, m_2\neq m_3)
\end{split}
\end{equation}

Although \emph{evaluation pushdown} and \emph{computation reuse} help reduce unnecessary computations, considerable computation is still required for evaluations involving the slowest and most expensive model $m^{*}$. In the following, we introduce techniques to accelerate this computation.

\begin{figure}[t]
    \centering
    \includegraphics[width=\linewidth]{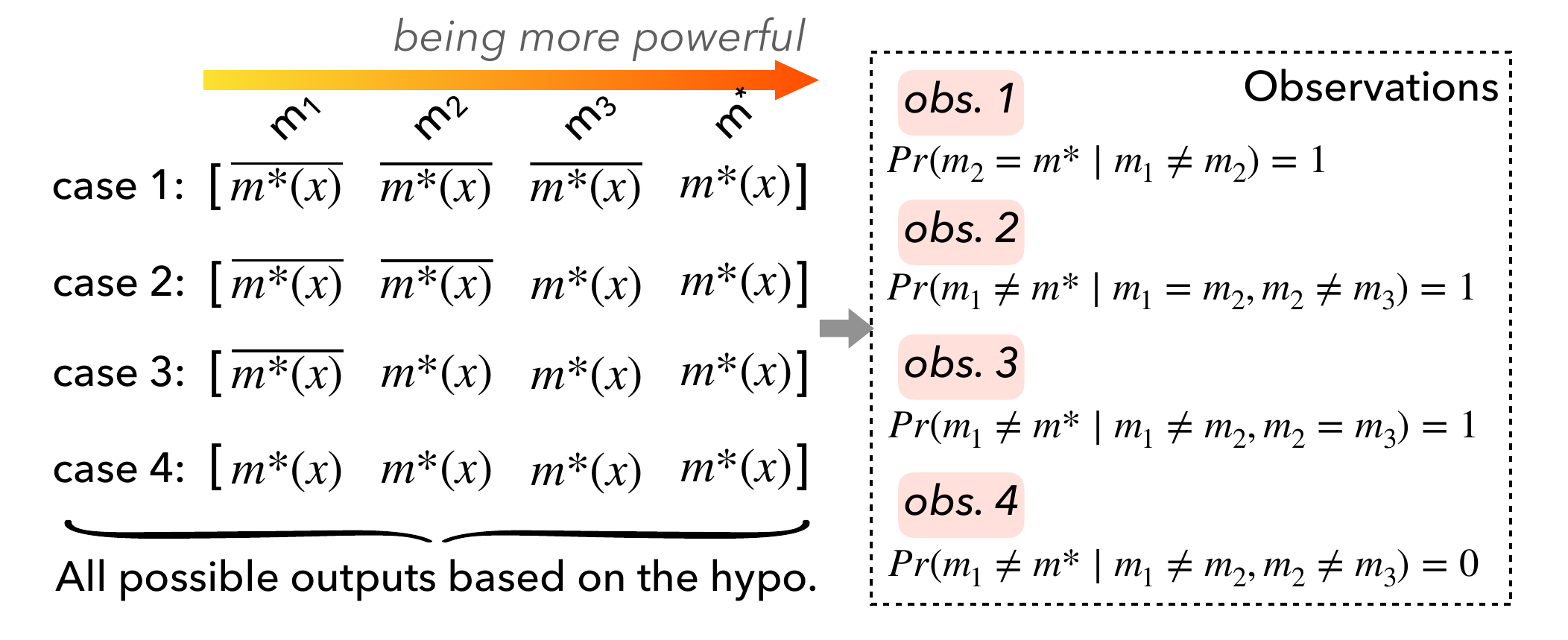}
    \caption{Possible responses of different models to a query, based on the model capability hypothesis. $m^{*}(x)$ denotes the response of the most powerful model $m^{*}$ to the input $x$, while $\overline{m^{*}(x)}$ denotes the opposite response to the same input.}
    \label{fig:hypothesis}
\end{figure}

\begin{table}[t] \small
    \centering
    \setlength{\abovecaptionskip}{1mm}
    \caption{Comparison of GPT-4.1 and GPT-4.1-nano outputs for the task of extracting amenities from estate descriptions.}
    \renewcommand{\arraystretch}{1.1}
    \setlength{\tabcolsep}{1mm}{
    \begin{tabular}{c c c c}
    \toprule[1pt]
    \#Aligned & \#Misaligned & \#GPT-4.1-is-right & \#GPT-4.1-nano-is-right \\ \hline
    424 & 76 & 69 & 7 \\
    \bottomrule[1pt]
    \end{tabular}}
    \label{tab:model_capability}
    \vspace{-3mm}
\end{table}

\subsection{Improvement Score Approximation}
To achieve acceleration, we first introduce a model capability hypothesis that enables the elimination of slow and expensive $m^{*}$ invocations when computing improvement scores.

\begin{hypothesis}[Model Capability Hypothesis]\label{hypo:model_capability}
    Given the ground-truth result $y$ to a query $x$, if a model $m$ returns a correct answer $y$ to $x$, \ie $m(x)=y$, any model $m^{+}$ that is more powerful than $m$ also returns the right answer $y$, \ie $m^{+}(x)=y$.
\end{hypothesis}

To showcase the generality of this hypothesis, we compare the outputs of GPT-4.1 and GPT-4.1-nano on the same task: extracting amenities from the detailed estate descriptions. We sample 500 records from the Estate dataset for evaluation. As shown in Table~\ref{tab:model_capability}, among the cases where the outputs of GPT-4.1 and GPT-4.1-nano differ, over 90\% of GPT-4.1 responses are correct.
Although the hypothesis do not hold universally, it is empirically valid in most practical cases.

Under this hypothesis, we can characterize the possible responses of models in the candidate set as illustrated on the left side of Figure~\ref{fig:hypothesis}, where we treat the output from the best model $m^{*}$ as the ground truth. In addition, several observations can be made on combinations of model outputs, as shown in Figure~\ref{fig:hypothesis}. For example, $m_2=m^{*}$ always holds when $m_1\neq m_2$ (observation 1 in Figure~\ref{fig:hypothesis}).

Using observation 1, we refine Eq.~\ref{eq:impr_compute12} and obtain the following approximation:
\begin{equation}\label{eq:impr_approx12}
\begin{split}
    I_{m_1\rightarrow m_2}&=Pr(m_2=m^{*}\mid m_1\neq m_2)Pr(m_1\neq m_2)\\
    &\approx Pr(m_1\neq m_2),
\end{split}
\end{equation}
where the evaluations of $m_2=m^{*}$ are eliminated completely thanks to the observations based on the hypothesis.

The same idea applies to $I_{m_1\rightarrow m_3}$. From Figure~\ref{fig:hypothesis}, we observe that $m_3=m^{*}$ always holds when both $m_2\neq m_3$ and $m_1=m_2$ are true. This leads to the following approximation of $I_{m_1\rightarrow m_3}$:
\begin{equation}\label{eq:impr_approx13}
    I_{m_1\rightarrow m_3}\approx I_{m_1\rightarrow m_2} + Pr(m_2\neq m_3\mid m_1=m_2)Pr(m_1=m_2)
\end{equation}
Note that since $Pr(m_1=m_2)=1-I_{m_1\rightarrow m_2}$, this computation can reuse the result of Eq.~\ref{eq:impr_approx12} so that $Pr(m_2\neq m_3\mid m_1=m_2)$ is the only term to be determined when computing $I_{m_1\rightarrow m_3}$. Compared to the full evaluation in Eq.~\ref{eq:impr_compute13}, the evaluations involving the expensive $m^{*}$ invocations are entirely eliminated in Eq.~\ref{eq:impr_approx13}, and the evaluation of $m_3$ is constrained to records satisfying $m_1 = m_2$, preserving the benefits of ``evaluation pushdown''.

Similarly, we can derive an approximation of $I_{m_1\rightarrow m^{*}}$ as follows. This result also follows observations from the hypothesis and is inferred by algebraic transformations (details are provided in Appendix~\ref{appx:proof} in the supplementary material).
\begin{equation}\label{eq:impr_approx14}
\begin{split}
    &I_{m_1\rightarrow m^{*}}\approx Pr(m_1\neq m^{*}\mid m_1=m_2,m_2=m_3)(1-I_{m_1\rightarrow m_3})+\\
    & I_{m_1\rightarrow m_3}-I_{m_1\rightarrow m_2}+Pr(m_2=m_3\mid m_1\neq m_2)Pr(m_1\neq m_2)
\end{split}
\end{equation}
This approximation reduces the number of invocations of $m^{*}$ considerably. Specifically, evaluations involving $m^{*}$ are required only on the subset of records that pass the conditions $m_1=m_2$ and $m_2=m_3$. Compared to Eq.~\ref{eq:impr_compute14}, the approximation reduces the computations by $3\times$ when computing $I_{m_1\rightarrow m^{*}}$.

\begin{table}[t] \small
    \centering
    \setlength{\abovecaptionskip}{1mm}
    \caption{Dataset characteristics.}
    \renewcommand{\arraystretch}{1.1}
    \setlength{\tabcolsep}{1mm}{
    \begin{tabular}{c|c|c|c}
    \toprule[1pt]
    Dataset & \#Records & \#Attributes & Modalities \\ \hline
    \textbf{Movie} & 250 & 22 & numerical values, text, images \\ \hline
    \textbf{Estate} & 1041 & 4 & images, long text \\ \hline
    \textbf{Game} & 18891 & 21 & dates, numerical values, images, text \\
    \bottomrule[1pt]
    \end{tabular}}
    \label{tab:dataset}
\end{table}

\section{Experimental Evaluation}\label{sec:experiment}
We evaluate \ours and several existing proposals on three real-world benchmarks. We also analyze systematically the advantages and limitations of design choices in \ours.

\subsection{Experimental Setup}
\subsubsection{Datasets} We use three well-established, publicly available multi-modal datasets for the evaluation; see Table~\ref{tab:dataset}.
(1) The \textbf{Movie}\footnote{\url{https://huggingface.co/datasets/moizmoizmoizmoiz/MovieRatingDB}} dataset includes 250 movie records sourced from OMDB, each with 22 attributes, including titles, directors, ratings, posters, and plot descriptions. It covers numerical, textual, and image modalities.
(2) The \textbf{Estate}\footnote{\url{https://huggingface.co/datasets/Binaryy/multimodal-real-estate-search}} dataset contains 1,041 estate entries, each with 4 attributes, including property images, project descriptions, locations, and estate details. The modalities include images and long text.
(3) The \textbf{Game}\footnote{\url{https://github.com/triesonyk/data-analysis-steam-games}} dataset contains 18,891 video game records with PEGI ratings from Steam. Each record has 21 attributes, including game titles, cover images, release dates, prices, and game summaries. The modalities span dates, numerical values, images, and long text.

\subsubsection{Workloads} For each dataset, we construct a query workload comprising 12 queries. These queries are categorized into three groups based on the number of operations involved:
(a) \textit{Small}: Each query contains a single operator. \textit{Small} queries usually process all records in a table, and they are too limited for optimization by traditional optimizers, whereas in \ours, they can be further optimized through non-LLM operator replacement and physical plan optimization.
(b) \textit{Medium}: Each query consists of two to three operators. \textit{Medium} queries present moderate optimization potential for both traditional systems and \ours.
(c) \textit{Large}: Each query has four or more operators. \textit{Large} queries provide increased opportunities for optimization, and they usually produce a single-value result, allowing us to evaluate result quality across systems.
The workloads are available in Appendix~\ref{appx:workloads} in the supplementary material.

\subsubsection{Systems} We compare \ours against three existing data analytics systems:
\begin{itemize}[topsep=0pt,leftmargin=*]
    \item \textsf{GPT-4.1}. A baseline that prompts the LLM directly by feeding the query in natural language and entire table into the model to obtain analytical results.
    \item \textsf{TableRAG}~\cite{chen2024tablerag}. A Retrieval-Augmented Generation (RAG) framework specifically designed for LM-based table understanding.
    \item \textsf{Table-LLaVA}~\cite{zheng2024tablellava}. A tabular multimodal large language model (MLLM), which treats tables as images and combines table images with user queries as LLM inputs. The model is deployed locally on an Nvidia A100 40G GPU.
    \item \textsf{Palimpzest}\footnote{\url{https://github.com/mitdbg/palimpzest/releases/tag/0.7.7}}. An LLM-powered analytical processing system that implements several operators to execute analytical queries. It employs a Cascade optimizer to enable logical and physical plan optimization tailored to semantic operators.
    \item \textsf{Lotus}\footnote{\url{https://github.com/lotus-data/lotus/releases/tag/v1.1.0}}. A Pandas-like programming framework with LLM-powered semantic operators, each of which can be executed using a standard algorithm or an approximate algorithm.
\end{itemize}

\subsubsection{Implementation} In the experiments, we employ the following OpenAI models as execution backends for semantic operators: \texttt{gpt-4.1-nano}, \texttt{gpt-4o-mini}, \texttt{gpt-4.1-mini}, \texttt{gpt-4.1}.The \texttt{gpt-4o} model is employed for logical plan optimization. Model pricing follows the official OpenAI pricing policy~\cite{pricing}. For \textsf{TableRAG}, we set the cell encoding budget to 10,000 and the retrieval limit 50. To mitigate the impact of LLMs varying in performance, we restrict the LLM choices of all systems to the GPT-4.1 family to ensure fair comparison. For logical plan optimization, the number of optimization iterations $N_{\mathit{max}}$ is set to $3$. The parameter $\lambda$ in Eq.~\ref{eq:sample} is set to 0.2. The error tolerance $\epsilon$ is set to $0.8$. For physical plan optimization, the improvement margin $\Delta I_{\mathit{min}}$ is set to $20\%$. The ratio of data samples is set to $5\%$. To improve throughput, \ours leverages concurrent programming with 16 coroutines at the most.

\begin{table*}[t] \small
    \centering
    \setlength{\abovecaptionskip}{1mm}
    \caption{Runtime and monetary cost of \ours, \textsf{Table-LLaVA}, \textsf{TableRAG}, \textsf{Palimpzest}, and \textsf{Lotus} on the Movie, Estate, and Game benchmarks. Mean values across queries are reported.}
    \renewcommand{\arraystretch}{1.1}
    \setlength{\tabcolsep}{1.7mm}{
    \begin{tabular}{c|c|c|c|c|c|c|c|c|c|c|>{\columncolor{light-gray}}c|>{\columncolor{light-gray}}c}
    \toprule[1pt]
    \multirow{2}{*}{Dataset} & \multirow{2}{*}{Workload} & \textsf{Table-LLaVA} & \multicolumn{2}{c|}{\textsf{TableRAG}} & \multicolumn{2}{c|}{\textsf{Palimpzest}} & \multicolumn{2}{c|}{\textsf{Lotus}} & \multicolumn{4}{c}{\ours} \\ \cline{3-13}
    & & Time (s) & Time (s) & Cost (\$) & Time (s) & Cost (\$) & Time (s) & Cost (\$) & Time (s) & Cost (\$) & $\Delta$ Time & $\Delta$ Cost \\ \hline
    \multirow{3}{*}{\textbf{Movie}} & \textit{Small} & $6.103$ & $18.048$ & $0.035$ & $81.754$ & $0.220$ & $40.283$ & $0.170$ & $19.963$ & $0.091$ & $\downarrow 50.44\%$ & $\downarrow 46.48\%$ \\
    & \textit{Medium} & $1.320$ & $15.452$ & $0.028$ & $82.685$ & $0.126$ & $62.326$ & $0.107$ & $16.956$ & $0.011$ & $\downarrow 72.79\%$ & $\downarrow 89.50\%$ \\
    & \textit{Large} & $1.596$ & $17.463$ & $0.086$ & $163.115$ & $0.390$ & $76.178$ & $0.383$ & $21.169$ & $0.014$ & $\downarrow 72.21\%$ & $\downarrow 96.31\%$ \\ \hline
    \multirow{3}{*}{\textbf{Estate}} & \textit{Small} & {/} & $22.178$ & $0.123$ &
    $363.671$ & $1.479$ & $149.945$ & $0.748$ & $114.196$ & $0.612$ & $\downarrow 23.84\%$ & $\downarrow 18.16\%$ \\
    & \textit{Medium} & {/} & $21.516$ & $0.153$ & $581.923$ & $1.660$ & $252.182$ & $1.527$ & $193.317$ & $0.524$ & $\downarrow 23.34\%$ & $\downarrow 65.90\%$ \\
    & \textit{Large} & {/} & $24.449$ & $0.146$ & $918.673$ & $2.677$ & $260.099$ & $1.564$ & $233.679$ & $0.417$ & $\downarrow 10.16\%$ & $\downarrow 73.37\%$ \\ \hline
    \multirow{3}{*}{\textbf{Game}} & \textit{Small} & {/} & $18.725$ & $0.035$ &
    $\ge 5.5$ hours & / & $1375.920$ & $4.116$ & $213.767$ & $0.084$ & $\downarrow 84.46\%$ & $\downarrow 97.96\%$ \\
    & \textit{Medium} & {/} & $16.485$ & $0.042$ & $\ge 5.5$ hours & / & $2491.667$ & $7.419$ & $478.463$ & $0.187$ & $\downarrow 80.80\%$ & $\downarrow 97.48\%$ \\
    & \textit{Large} & {/} & $15.961$ & $0.047$ & $\ge 5.5$ hours & / & $3664.475$ & $10.925$ & $728.443$ & $0.249$ & $\downarrow 80.12\%$ & $\downarrow 97.72\%$ \\ 
    \bottomrule[1pt]
    \end{tabular}}
    \label{tab:compare}
\end{table*}

\begin{table}[t] \small
    \centering
    \setlength{\abovecaptionskip}{1mm}
    \caption{Quality evaluation of \ours, its competitors, and its variants across the three benchmarks.}
    \renewcommand{\arraystretch}{1.1}
    \begin{tabular}{p{4cm} c c c}
    \toprule[1pt]
     & \textbf{Movie} & \textbf{Estate} & \textbf{Game} \\ \hline
     \textsf{GPT-4.1} & \ding{55} & \ding{55} & \ding{55} \\
     \textsf{Table-LLaVA} & 0\% & \ding{55} & \ding{55} \\
     \textsf{TableRAG} & 0\% & 0\% & 0\% \\
     \textsf{Palimpzest} & 58.3\% & 25.0\% & / \\
     \textsf{Lotus} & 8.3\% & 16.7\% & 25.0\% \\ \hline
     \ours & 75.0\% & 41.7\% & 50.0\% \\
     \textsf{\ours w/o Logical optim.} & 58.3\% & 33.3\% & 41.7\% \\
     \textsf{\ours w/ GPT-4.1} & 58.3\% & 33.3\% & 33.3\% \\
     \textsf{\ours w/o all optim.} & 66.7\% & 41.7\% & 41.7\% \\
    \bottomrule[1pt]
    \end{tabular}
    \label{tab:quality}
\end{table}

\subsection{Overall Results}
\subsubsection{Settings} We compare \ours to two state-of-the-art bulk semantic processing systems, \textsf{Palimpzest} and \textsf{Lotus}, in terms of runtime, monetary cost, and analytical quality. For fair comparison, \textsf{Palimpzest} is configured with access to \texttt{gpt-4.1}, \texttt{gpt-4.1-nano}, \texttt{text-embedding-3-small}, and \texttt{clip-ViT-B-32}. All built-in implementation rules in \textsf{Palimpzest} are enabled, and its optimization policy is set to \texttt{MinTime} on Movie and Estate and \texttt{MinCost} on Game. For \textsf{Lotus}, we use \texttt{gpt-4.1} as the backend model and \texttt{gpt-4.1-nano} as the helper model. Both \textsf{Palimpzest} and \textsf{Lotus} are configured to use concurrent programming, with the number of workers set to 16.

\subsubsection{System runtime} Table~\ref{tab:compare} presents runtime and monetary cost comparisons across all three datasets. \ours consistently outperforms \textsf{Palimpzest} and \textsf{Lotus} in runtime, reducing the execution time by 65.15\%, 19.11\%, and 81.87\% on Movie, Estate, and Game, respectively. Notably, the runtime of \textsf{Palimpzest} on Game exceeds 5.5 hours, making it impractical for real-world deployment. The runtime improvements are attributed to both the logical and the physical optimizations in \ours. Compared to \textsf{Lotus}, which supports proxy inference using smaller models, \ours additionally reduces LLM computations through operator merging, reordering, and replacement. While \textsf{Palimpzest} also supports operator reordering, \ours further enhances performance through collaborative execution between large and small models. Although \textsf{Table-LLaVA} performs tableQA efficiently (completing queries in just a few seconds), the image size can easily exceed the input limit as table length grows. \textsf{TableRAG} achieves nearly constant runtime regardless of table size because it retrieves only a fixed number of relevant records. However, as discussed later, both \textsf{Table-LLaVA} and \textsf{TableRAG} exhibit extremely poor answer quality.

\subsubsection{System cost} In terms of monetary cost, \ours reduces monetary cost significantly compared to \textsf{Palimpzest} and \textsf{Lotus}. The reasons are two-fold: First, the three transformation rules in our logical plan optimizer effectively reduce LLM invocations. Second, the physical optimizer assigns cheaper, smaller models where appropriate. Since this cost reduction applies at the per-record level, the more data is processed, the larger the savings.

\subsubsection{Result quality} Table~\ref{tab:quality} shows the analytical accuracy of \ours and the five other systems on the three datasets. The quality is measured as the proportion of queries answered correctly, where the ground truth of each query is labeled manually. The LLM, GPT-4.1, fails to answer any query across the datasets because the table content exceeds the maximum context length. Even on a smaller version of the smallest dataset, Movie, where only 5 out of 22 columns are selected, it fails. \textsf{Table-LLaVA} also fails to produce correct answers, as the image quality is inadequate even on Movie. For Estate (and also Game), the generated image sizes (245,668,890 px) exceed the model’s limit of 178,956,970 px. \textsf{TableRAG} achieves an accuracy of 0 because it cannot process data outside its retrieval scope and does not support aggregation operations. \ours achieves better accuracy than \textsf{Palimpzest} and \textsf{Lotus} across most datasets. However, all systems struggle on queries involving relatively complex mathematical computations over string-like values, such as calculating the average price of a large group of properties.

\begin{figure*}[t]
    \centering
    \includegraphics[width=\textwidth]{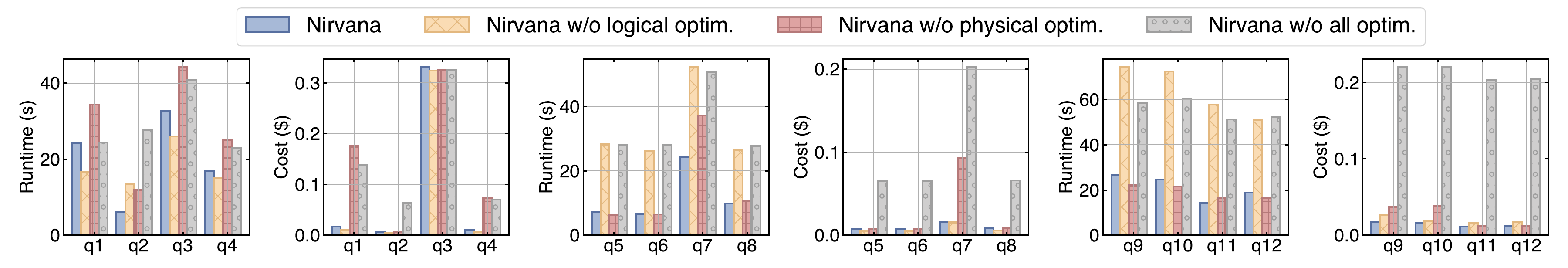}
    \vspace{-8mm}
    \caption{Ablation study of \ours in terms of runtimes and monetary costs on Movie.}
    \label{fig:RQ2-movie}
    \vspace{-3mm}
\end{figure*}
\begin{figure*}[t]
    \centering
    \includegraphics[width=\textwidth]{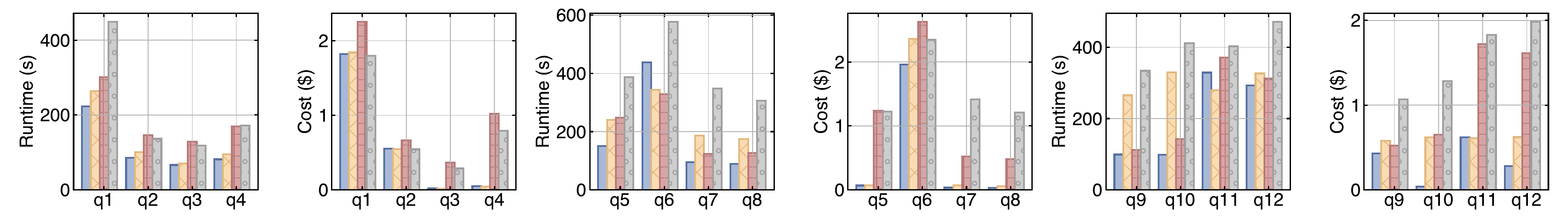}
    \vspace{-8mm}
    \caption{Ablation study of \ours in terms of runtimes and monetary costs on Movie on Estate.}
    \label{fig:RQ2-estate}
    \vspace{-3mm}
\end{figure*}
\begin{figure*}[t]
    \centering
    \setlength{\belowcaptionskip}{-3mm}
    \includegraphics[width=\textwidth]{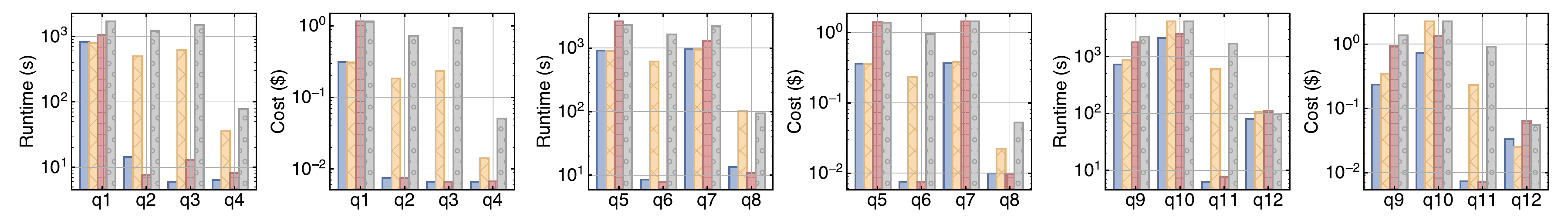}
    \vspace{-8mm}
    \caption{Ablation study of \ours in terms of runtime and monetary costs on Game. The $y$-axes use a logarithmic scale.}
    \label{fig:RQ2-steam}
\end{figure*}

\begin{table}[t] \small
    \centering
    \setlength{\abovecaptionskip}{1mm}
    \caption{Comparison w.r.t. cost of processing query \sql{q10} on the Estate dataset between \ours's logical optimizer and the Cascades optimizer in \textsf{Palimpzest (PZ)}. Times are in seconds; monetary costs are in USD.}
    \renewcommand{\arraystretch}{1.1}
    \begin{tabular}{p{2cm} c c c c}
    \toprule[1pt]
     & Opt. time & Opt. cost & Exec. time & Exec. cost \\ \hline
     \textsf{PZ (Cascades)} & $\sim0.0$ & $0.0$ & $995.3$ & $2.4$ \\
     \ours & $9.8$ & $8.2E^{-3}$ & $99.1$ & $0.038$ \\
    \bottomrule[1pt]
    \end{tabular}
    \label{tab:logical_optim}
\end{table}

\subsection{Logical Plan Optimizer Evaluation}\label{sec:logical_optim}
\subsubsection{Settings} We perform an ablation analysis to evaluate the effectiveness of \ours's logical optimizer. Two variants are considered: (1) \textsf{\ours w/o logical optim.} that disables the logical plan optimizer; and (2) \textsf{\ours w/o all optim.} that disables both the logical optimizer and physical optimizer.

\subsubsection{Runtime and cost} Figures~\ref{fig:RQ2-movie},~\ref{fig:RQ2-estate}, and~\ref{fig:RQ2-steam} show runtime and monetary costs for \ours and its variants across all queries. Generally speaking, across almost all datasets and workloads, both the logical and physical optimizers independently contribute to performance gains, reducing runtime and cost compared to the variant with no optimizations. Thus, logical and physical plan optimizers play their unique roles in query optimization, reducing both runtime and cost considerably. Runtime and cost differences become more pronounced for larger workloads (\eg queries q9–q12) and larger datasets (\eg Game), where optimization opportunities increase. Comparing \ours to \textsf{\ours w/o logical optim.}, we observe that the logical optimizer accelerates query execution and lowers cost notably in most cases, especially for ``large'' queries (q5–q12), where optimizations like filter pushdown effectively prune irrelevant data early. For smaller workloads (q1–q4), the benefits of logical optimizations are minimal, as these queries usually involve only a single operator. In these cases, the logical plan optimizer in \ours has little to do given a single operation, except applying transformation like non-LLM replacement. Indeed, only a few queries --- q2 on both Movie and Game --- qualify for such replacement. Interestingly, in some cases (\eg q1 and q3 in Figure~\ref{fig:RQ2-movie}, and q1 in Figure~\ref{fig:RQ2-steam}), \ours incurs higher runtime and cost than the variant without logical optimization for analytical processing. This is due to the high runtime and cost of the logical plan optimizer that invokes multiple LLMs during the optimization process.

\subsubsection{Comparison with \textsf{Palimpzest}} We report the overhead of the logical optimizer in Table~\ref{tab:logical_optim} as a case study. We see that, processing query q10 on Estate requires 9.8 seconds and 0.82 cents for logical optimization alone, compared to the zero-cost, near-instant optimization performed by the Cascades optimizer in \textsf{Palimpzest}. Reducing this overhead presents an important direction for future research on LLM-powered query optimization.

\subsubsection{Result quality} We assess the impact of the optimizers on result quality using the results in Table~\ref{tab:quality}. \ours with full optimization achieves the same number of correct answers as its no-optimization variant (\textsf{\ours w/o all optim.}), while having considerably lower runtime and cost. However, disabling logical optimization (\ie \textsf{\ours w/o logical optim.}) decreases accuracy slightly. This is likely due to a higher likelihood of inference errors when more records are processed by a weaker model, as fewer filters are applied early. And also, an LLM might make more mistakes than a computing function if non-LLM replacement is not enabled. However, the difference is marginal with typically only one or two additional incorrect results out of 12 results if only using either logical or physical plan optimizer.

\subsubsection{Reliability of the LLM-as-a-Judge} We collect all query rewrites and their rating scores assigned by the LLM-as-a-Judge from the optimization process. We categorize these rewrites into four groups: (a) True Positives (TP): correct rewrite \& rating $\ge\epsilon$ (0.8); (b) False Positives (FP): incorrect rewrite \& high rating; (c) False Negatives (FN): correct rewrite \& low rating; and (d) True Negatives (TN): incorrect rewrite \& low rating. In the evaluation, the correctness of a query rewrite is determined by a human expert. We use the following metrics to measure the reliability of the LLM-as-a-Judge: (1) Success rate: $\frac{\mathit{\#TP}+\mathit{\#TN}}{\mathit{\#rewrites}}$; (2) Precision: $\frac{\mathit{\#TP}}{\mathit{\#TP}+\mathit{\#FP}}$; (3) Recall: $\frac{\mathit{\#TP}}{\mathit{\#TP}+\mathit{\#FN}}$.
Table~\ref{tab:llm-as-judge} reports the results and the cost of verification.
The LLM-as-a-Judge achieves $>80\%$ success rate, $>85\%$ precision, and $>90\%$ recall across three benchmarks. These results demonstrate near-human verification quality at a reasonable cost. 
We also note that failures occur in $\sim10\%$ of cases (failing to abort incorrect rewrites) and $\sim8\%$ of cases (failing to accept correct rewrites), mainly due to low coverage of data samples for validation and heterogeneous operator outputs that are too vague to verify.

\subsubsection{Optimizer w/ semantics v.s. optimizer w/o semantics} To assess the effectiveness of semantics in logical optimization, we conduct ablation studies to isolate the impact of semantic-aware transformations (\ie non-LLM replacement v.s. filter pushdown and operator fusion). Table~\ref{tab:sem_in_logic} reports the average runtimes and costs of both optimization and end-to-end execution across workloads. incorporating semantic-aware transformations consistently reduces overall system runtime and cost, albeit with minor optimization overhead, particularly on Movie, where many user-defined operations can be implemented directly as UDFs.
In addition, we compare our optimizer with a ``two-step'' hybrid optimization that first applies basic transformations (\ie filter pushdown and operator fusion) and then applies semantic-aware ones(\ie Non-LLM replacement). As shown, the total costs and end-to-end runtimes of the two approaches are nearly identical.

\begin{table}[t] \small
    \centering
    \setlength{\abovecaptionskip}{1mm}
    \caption{Reliability of the LLM-as-a-Judge and its average cost across queries.}
    \renewcommand{\arraystretch}{1.1}
    \begin{tabular}{p{2cm} c c c c}
    \toprule[1pt]
     & Success rate & Precision & Recall & Cost \\ \hline
     \textbf{Movie} & $81.6\%$ & $85.2\%$ & $93.8\%$ & $0.0024$ \\
     \textbf{Estate} & $90.0\%$ & $95.6\%$ & $91.6\%$ & $0.0026$ \\
     \textbf{Game} & $86.7\%$ & $84.6\%$ & $100\%$ & $0.0016$ \\
    \bottomrule[1pt]
    \end{tabular}
    \label{tab:llm-as-judge}
\end{table}

\begin{table}[t] \small
    \centering
    \setlength{\abovecaptionskip}{1mm}
    \caption{Logical optimizer w/ semantics v.s. w/o semantics. Times are in seconds and costs are in $\times10^{-2}$ USD.}
    \renewcommand{\arraystretch}{1.1}
    \begin{tabular}{l c c c c c c}
    \toprule[1pt]
     & \multicolumn{3}{c}{\textbf{Movie}} & \multicolumn{3}{c}{\textbf{Estate}} \\
     & S & M & L & S & M & L \\ \hline
     \textsf{Optim. time (w/ sem)} & 6.57 & 7.32 & 8.20 & 7.45 & 8.88 & 9.49 \\
     \textsf{Optim. time (w/o sem)} & 6.81 & 7.49 & 8.05 & 7.03 & 7.87 & 8.76 \\
     \textsf{Optim. time (2 step)} & $6.54$ & $7.15$ & $8.20$ & 7.44 & 8.85 & 8.17 \\
     \textsf{Overall time (w/ sem)} & 19.96 & 12.06 & 21.17 & 114.19 & 193.32 & 205.08 \\
     \textsf{Overall time (w/o sem)} & 34.13 & 32.95 & 50.82 & 101.71 & 192.28 & 227.92 \\
     \textsf{Overall time (2 step)} & 19.96 & 12.06 & 21.17 & 114.19 & 193.32 & 205.08 \\ \hline
     \textsf{Optim. cost (w/ sem)} & 0.67 & 0.78 & 0.83 & 0.63 & 0.72 & 0.89 \\
     \textsf{Optim. cost (w/o sem)} & 0.53 & 0.61 & 0.82 & 0.57 & 0.65 & 0.69 \\
     \textsf{Optim. cost (2 step)} & 0.52 & 0.60 & 0.81 & 0.55 & 0.64 & 0.67 \\
     \textsf{Overall cost (w/ sem)} & 9.14 & 1.00 & 1.41 & 61.19 & 52.39 & 34.21 \\
     \textsf{Overall cost (w/o sem)} & 9.21 & 1.22 & 2.00 & 61.00 & 51.69 & 44.69 \\
     \textsf{Overall cost (2 step)} & 9.13 & 0.99 & 1.40 & 61.17 & 52.38 & 34.19 \\
    \bottomrule[1pt]
    \end{tabular}
    \label{tab:sem_in_logic}
\end{table}

\begin{table}[t] \small
    \centering
    \setlength{\abovecaptionskip}{1mm}
    \caption{Overhead comparison of processing \sql{q3} on Estate using \textsf{Smart} and the physical optimization in \ours.}
    \renewcommand{\arraystretch}{1.1}
    \begin{tabular}{p{3cm} c c c}
    \toprule[1pt]
     & Optim. time (s) & Exec. time (s) & Ratio \\ \hline
     \textsf{Smart (exhaustive)} & 59.06 & 626.77 & 9.42\% \\
     \textsf{Smart (efficient)} & 53.80 & 620.95 & 8.66\% \\
     \textsf{Smart (multi-model)} & 56.27 & 625.73 & 8.99\% \\ \hline
     \textsf{\ours (Synchronous)} & 13.11 & 674.56 & 1.94\% \\
     \textsf{\ours (Asynchronous)} & 4.12 & 66.47 & 5.82\% \\
    \bottomrule[1pt]
    \end{tabular}
    \label{tab:physical_optim}
\end{table}

\subsection{Physical Plan Optimizer Evaluation}
\subsubsection{Settings} We conduct an ablation study of the physical optimizer, comparing \ours with two variants: (1) \textsf{\ours w/o physical optim.}, which disables the physical optimizer, and (2) \textsf{\ours w/o all optim.}, which disables both optimizers.

\subsubsection{Runtime and cost} The physical optimizer provides a robust strategy for reducing runtime and monetary cost. \textsf{\ours w/o physical optim.} outperforms the version with no optimization, offering evidence of the physical optimizer's effectiveness. Further, we observe that \textsf{\ours w/o physical optim.} often incurs lower costs than \textsf{\ours w/o logical optim.} We conclude that the physical optimizer plays a dominant role in cost reduction. Interestingly, we also find that in certain queries --- such as q5 and q6 in Figure~\ref{fig:RQ2-movie} --- logical plan optimization alone yields larger runtime and cost savings than when using also physical plan optimization. Another interesting finding is that logical optimization can bring more reductions in runtime and cost than physical plan optimization in certain cases where a semantic operator can be replaced by a computing operator (\eg q5 and q6 in Figure~\ref{fig:RQ2-movie}). This happens when semantic operators are replaced by traditional computing operators, showcasing that the logical optimizer can sometimes apply more aggressive transformations that yield larger runtime reductions.

\subsubsection{Result quality} We evaluate the effect of optimizers on result quality; see Table~\ref{tab:quality}. The findings align with the findings in Section~\ref{sec:logical_optim}: the impact of optimization on answer quality is minimal.


\subsubsection{Comparison with Smart~\cite{jo2024smart}} Model selection is at the core of physical plan optimization. \textsf{Smart}~\cite{jo2024smart} is a recent system that proposes a cost-aware model selection algorithm with accuracy guarantees. However, it focuses on a single \sql{filter} operator and executes in a non-parallel fashion. It offers three variants: \textsf{exhaustive}, \textsf{efficient}, and \textsf{multi-model}. We compare \ours's physical optimizer in both its non-parallel (\textsf{Synchronous}) and parallel (\textsf{Asynchronous}) modes against all three \textsf{Smart} variants where the candidate LLMs are same as those available to \ours. Results are shown in Table~\ref{tab:physical_optim}. First, in the synchronous setting, \ours and \textsf{Smart} yield comparable execution times, suggesting that both systems make similar model selection decisions during query planning (\ie assigning GPT-4.1-nano to all operators as the execution backend by both Smart and our physical optimizer). However, \ours's optimizer requires significantly less time for optimization --- about one-fourth that of \textsf{Smart}. In the asynchronous setting, \ours drastically reduces both optimization and execution time, showing that parallel execution can boost throughput.


\begin{figure}[t]
    \centering
    \includegraphics[width=3.3in]{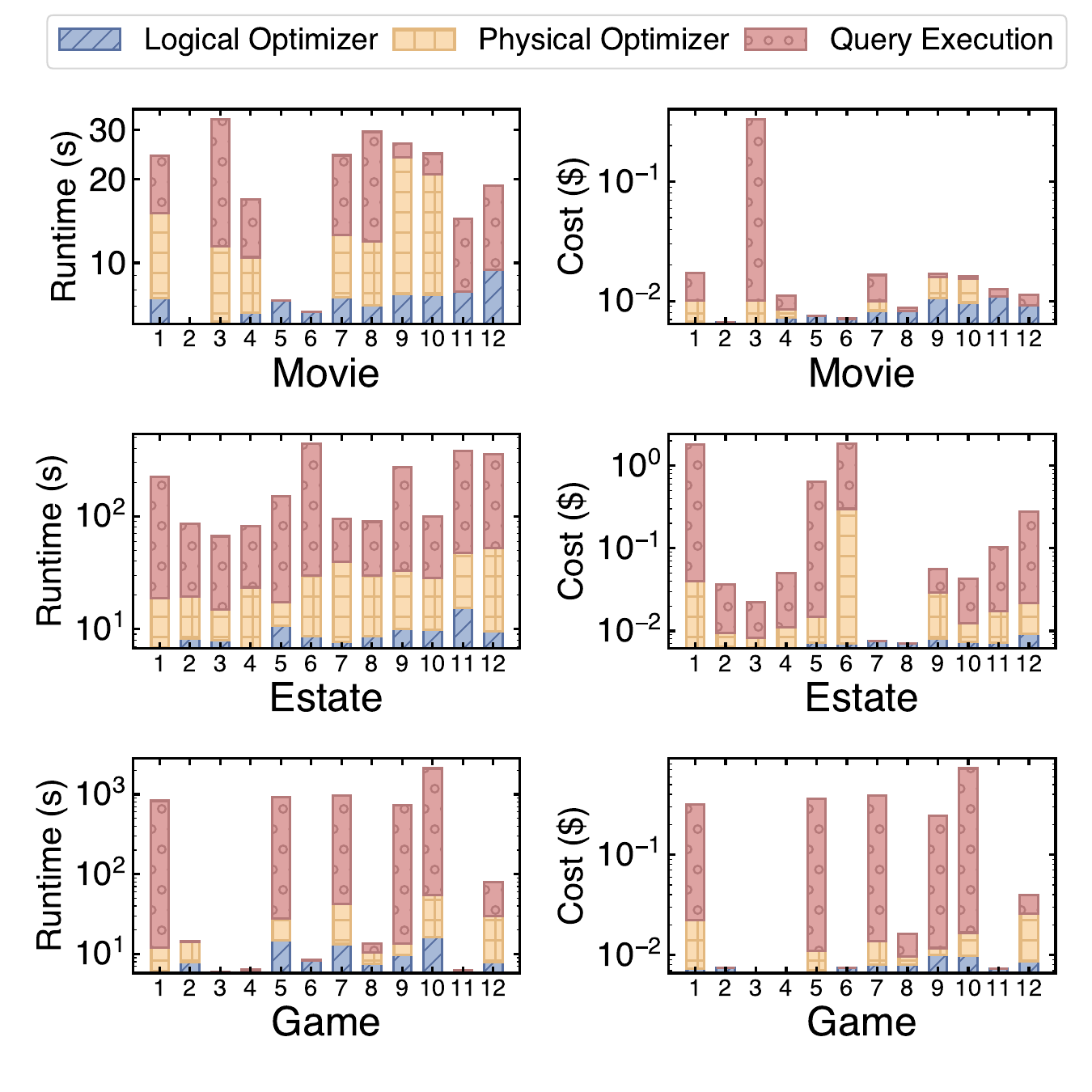}
    \vspace{-3mm}
    \caption{Breakdown of \ours in terms of the runtime and the cost per phase. The $y$-axes use a logarithmic scale.}
    \label{fig:cost-breakdown}
    \vspace{-3mm}
\end{figure}

\begin{figure}[t]
    \centering
    \includegraphics[width=3.3in]{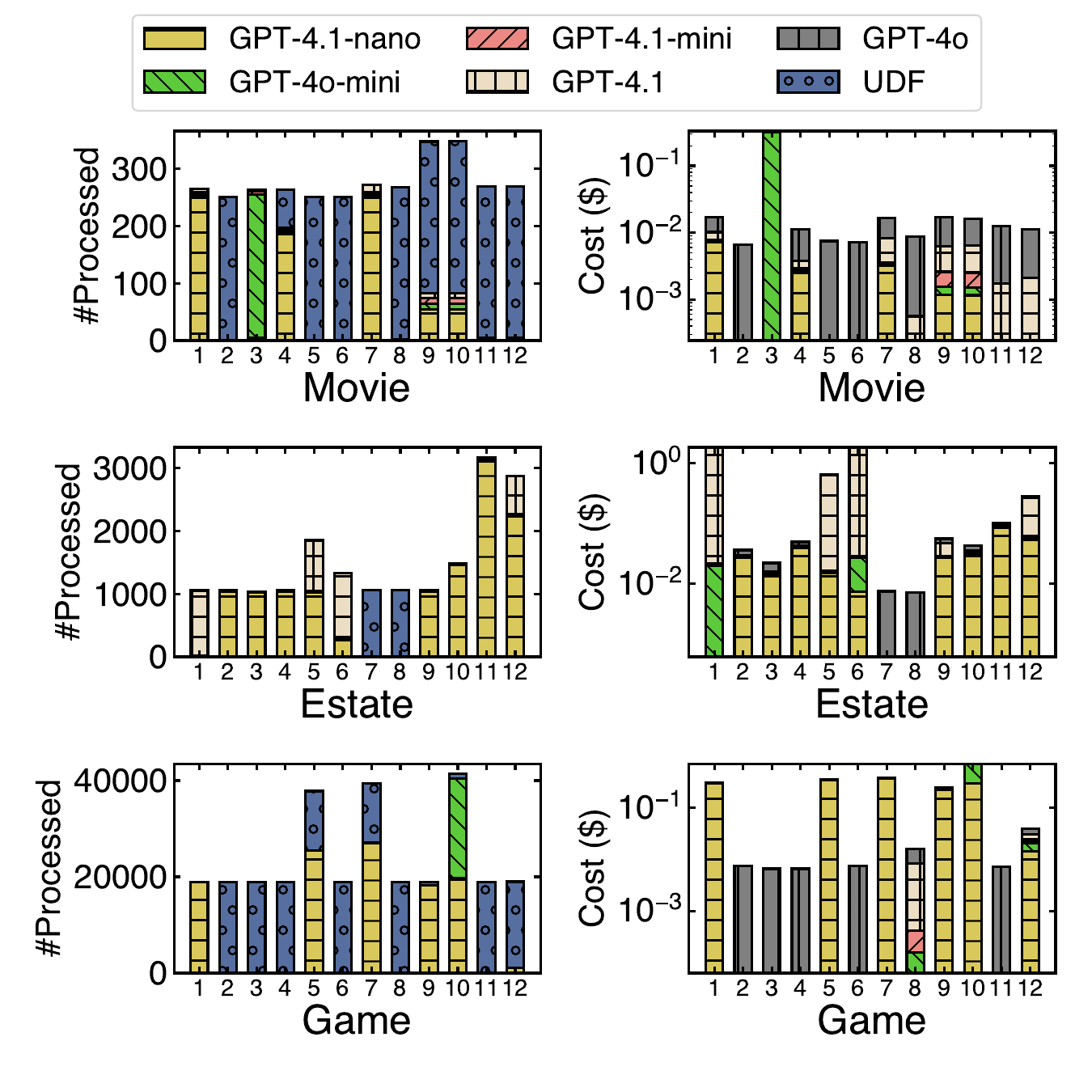}
    \vspace{-3mm}
    \caption{Breakdown of \ours in terms of the number of records processed and prices per models on all workloads.}
    \label{fig:model-breakdown}
    \vspace{-3mm}
\end{figure}

\begin{figure}[t]
    \centering
    \includegraphics[width=\linewidth]{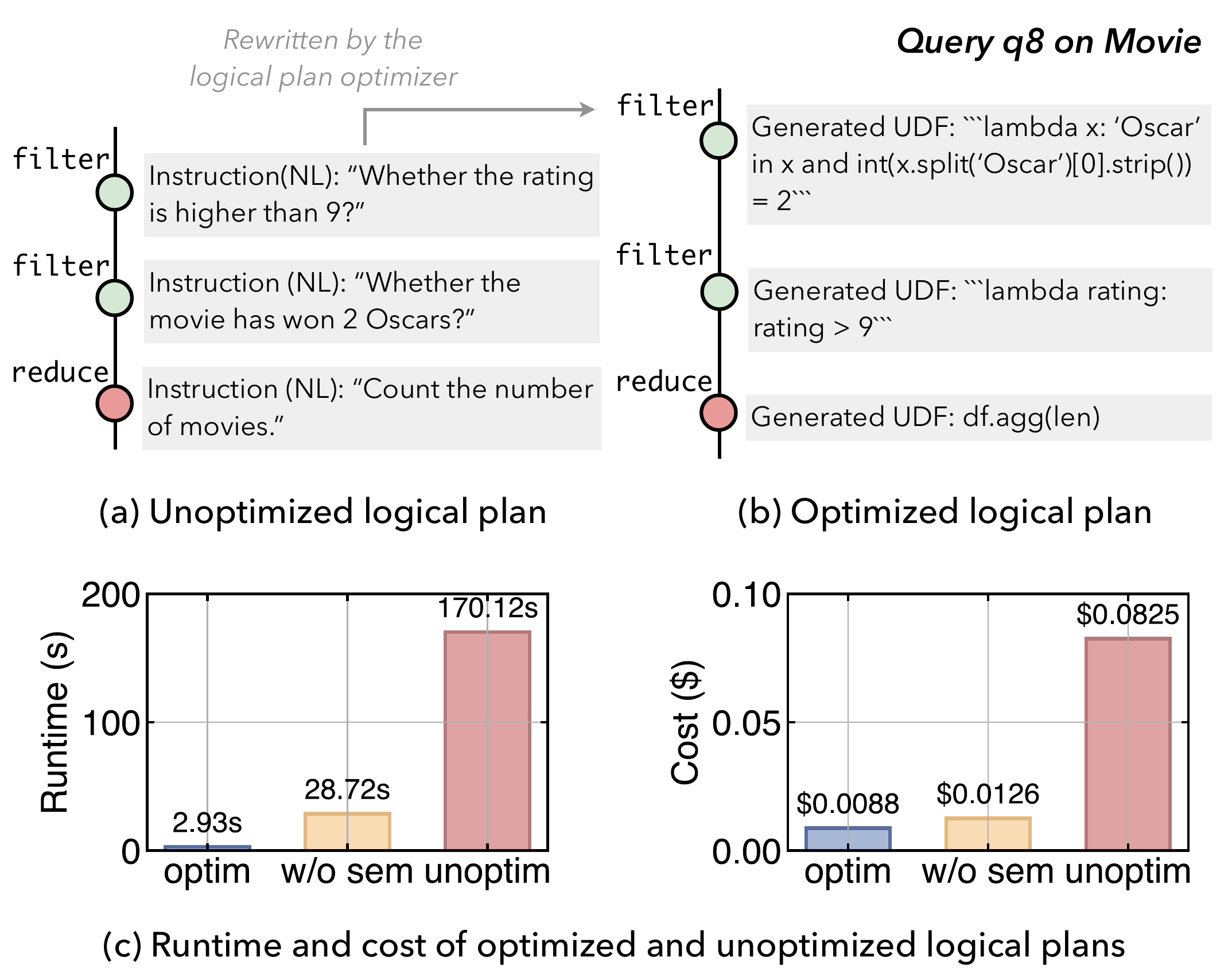}
    \vspace{-3mm}
    \caption{Case study of query optimization for \sql{q8} on Movie. The system runtime and cost when adopting the optimizer, the optimizer w/o semantics, and no optimizer are reported.}
    \label{fig:RQ2-logical-case}
    \vspace{-3mm}
\end{figure}

\subsection{Breakdown Analysis}
\subsubsection{Cost breakdowns} We provide detailed cost breakdowns of \ours (including both runtimes and API expenses) on all workloads. Figure~\ref{fig:cost-breakdown} shows the runtime and cost distributions across the three main components of \ours: the logical optimizer, the physical optimizer, and the query executor.
The proportions of optimization time relative to total query evaluation time are 50.7\%, 6.7\%, and 42.7\% for the Movie, Estate, and Game benchmarks, respectively.
By analyzing individual queries, we find that the impact of optimization depends on query execution: for queries using LLM-executed data operations (\eg Estate), the time taken on optimization is a small fraction of total processing time; for queries executed via UDFs generated by the optimizer (\eg \sql{q5}, \sql{q6} in Movie; \sql{q2}, \sql{q8} in Game), optimization becomes the dominant factor in query processing.
Furthermore, the ablation studies in Figures~\ref{fig:RQ2-movie},~\ref{fig:RQ2-estate}, and~\ref{fig:RQ2-steam} show that optimization significantly reduces end-to-end runtime costs (\eg runtimes are reduced by 53.7\%, 23.5\%, and 44.5\% on three datasets).
Another notable observation is that the runtime overhead of the logical optimizer remains nearly constant across datasets, incurring only 7--9 seconds on average (Movie: 7.31s, Estate: 9.17s, Game: 9.17s). This stability arises because the logical optimization cost depends primarily on the number of operators in a query rather than the dataset size. These results demonstrate that the benefits of LLM-based optimization outweigh the additional overhead.

\subsubsection{Model breakdowns} Figure~\ref{fig:model-breakdown} provides a detailed view of \ours in terms of the number of records processed by each model/executor and their corresponding costs. As observed, with the physical optimizer, \ours seamlessly transitions to utilizing LLMs with varying intelligence and prices while keeping a high accuracy. For instance, on \sql{q1} in Movie, GPT-4.1-nano is responsible for the majority of processing instead of the default GPT-4.1 model, reducing a great deal of cost. Moreover, different LLMs are selectively assigned to different operators within a query to balance accuracy and efficiency. For instance, on \sql{q8} in Game, \ours employs a combination of GPT-4.1-nano, GPT-4.1-mini, GPT-4o-mini, and GPT-4.1 for query processing.

\begin{figure}[t]
    \centering
    \includegraphics[width=\linewidth]{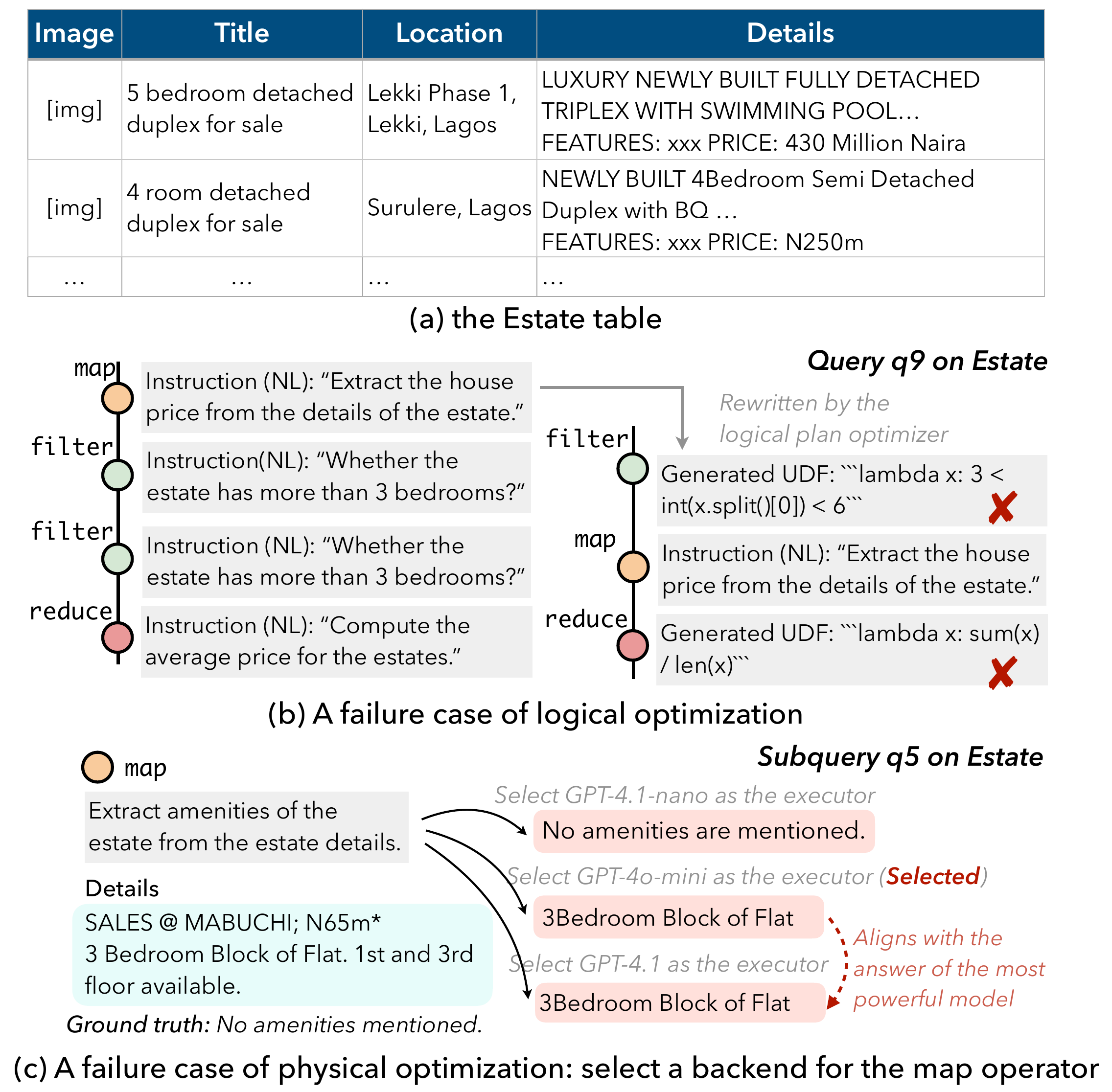}
    \vspace{-6mm}
    \caption{Failure cases of logical and physical optimization.}
    \label{fig:failure-case}
    \vspace{-5mm}
\end{figure}

\subsection{Case Studies}

\subsubsection{Successful cases} Figure~\ref{fig:RQ2-logical-case} illustrates the case of processing \sql{q8} on Movie. We notice that the logical optimizer uses two transformation rules to rewrite the initial logical plan: (a) it randomly reorders two \sql{filter} operators, as the optimizer has no knowledge of their selectivities; and (b) it substitutes the LLM-powered operations with non-LLM operations executed by computing functions. The LLM leverages query semantics to perform complex query rewrites, \eg converting the instruction ``whether the movie has ever won 2 Oscar'' into a UDF \sql{lambda x: `Oscar' in x and int(x.split(`Oscar')[0].strip())==2}. With optimizations, the runtime and cost are reduced by approximately 98\% and 89\%, respectively.

\subsubsection{Failure cases} Figure~\ref{fig:failure-case} illustrates representative failure cases in the logical and physical plan optimization. As shown in Figure~\ref{fig:failure-case}(b), the optimizer generates incorrect UDFs for the \sql{map} and \sql{reduce} operations. Specifically, the \sql{map} function works correctly only on a subset of the data records, while \sql{reduce} cannot process string values. This failure occurs because the optimizer lacks knowledge of the table contents to guide plan rewrites, and the LLM-as-a-Judge fails to detect and abort incorrect rewrites due to insufficient coverage of sampled data. Figure~\ref{fig:failure-case}(c) demonstrates a failure in physical plan optimization, where GPT-4o-mini is mistakenly assigned to an operator. This occurs because GPT-4o-mini produces the same incorrect output as the most powerful model, GPT-4.1, which is used as a proxy for the unknown ground truth.

\section{Related Work}\label{sec:related_work}
We proceed to review related work and explain how \ours differs from existing approaches.

\noindent\textbf{AI-augmented analytical processing systems}. The most relevant line of work involves building semantic data analytical processing systems from scratch, where semantic operators are powered by LLMs~\cite{patel2024lotus, liu2024palimpzest, cheng2023binder, madden2024unbound, shankar2024docetl, anderson2024aryn, lin2024zendb}. Representative examples include \textsf{Lotus}~\cite{patel2024lotus} and \textsf{Palimpzest}~\cite{liu2024palimpzest}, which share similar motivations with \ours. Lotus adopts an optimization strategy that selects an appropriate execution algorithm for each operator under accuracy constraints. Palimpzest introduces the Abacus optimizer~\cite{russo2025abacus}, inspired by the Cascades optimizer~\cite{gradfe1995cascades} from relational systems. However, its plan cost estimation and analysis quality evaluation remain coarse-grained. Compared to them, \ours attempts to explore conceptually distinct optimization strategies and discuss their pros and cons through experimental evaluations.

Another line of research focuses on integrating AI-powered semantic operators into mature relational DBMSs~\cite{jo2024thalamusdb, urban2024caesura, mindsdb}. For example, ThalamusDB~\cite{jo2024thalamusdb} adds a semantic filter operator powered by pretrained models for unstructured data (\eg text, images, audio) into DuckDB. CAESURA~\cite{urban2024caesura} augments traditional SQL operators (\eg joins, aggregations) with semantic operators such as TextQA and VisualQA to support reasoning over text and images. A key challenge is the difficulty of co-optimizing relational and semantic operators within conventional relational architectures.


\noindent\textbf{Workflow generation by LLMs}. Workflow generation for reasoning tasks via LLMs is conceptually related to logical plan optimization. Aflow~\cite{zhang2025aflow} samples reasoning workflows from scratch to achieve high accuracy on tasks such as math and commonsense QA. ADAS~\cite{hu2024adas} treats workflow generation as code generation and leverages LLMs to generate reasoning programs. GPTSwarm~\cite{zhuge2024gptswarm} models workflow generation as computation graph construction and proposes a graph generation strategy. Unlike these systems, \ours focuses on optimizing logical plans for semantic data analytical processing. It draws from traditional database principles such as rule-based transformations and cost-based optimization, while balancing analysis quality and system cost.

\noindent\textbf{Optimizations for compound AI systems}. Several recent efforts aim to optimize compound AI systems using techniques such as model selection and query reordering. QuestCache~\cite{liu2025optimizing} reduces LLM invocation costs for semantic data analytics by reordering table rows and columns to improve KV-cache reuse. Smart~\cite{jo2024smart} reduces the cost of semantic processing by selectively replacing powerful LLMs with smaller, cheaper models while maintaining similar output quality. While sharing a similar goal, \ours's physical plan optimizer adopts a distinct strategy. It introduces a novel cost-effectiveness metric and incorporates optimization techniques to reduce both runtime and monetary cost.
\section{Conclusion}\label{sec:conclusion}
With the goal of enabling semantic analytical processing over multi-modal data, we present \ours, a new LLM-powered semantic data analytics system equipped with semantic operators. To minimize system costs, \ours incorporates two novel query optimization techniques: an agentic logical plan optimizer based on random walk search, and a physical plan optimizer that allocates cost-effective models to operators in a query plan. Extensive experiments show that \ours is able to consistently outperform existing semantic data analytics systems in terms of both runtime and monetary cost.


\balance

\bibliographystyle{ACM-Reference-Format}
\bibliography{ref}

\newpage
\onecolumn
\appendix
\section{Proof of Eq.~\ref{eq:impr_approx14}}\label{appx:proof}
Starting from Eq.~\ref{eq:impr_compute14}, we first reformulate the first term after the equal sign:
\begin{equation*}
\begin{split}
    Pr(m_1\neq m^{*},m_1=m_2,m_2=m_3)&=Pr(m_1\neq m^{*}\mid m_1=m_2,m_2=m_3)Pr(m_2=m_3\mid m_1=m_2)Pr(m_1=m_2)\\
    &=Pr(m_1\neq m^{*}\mid m_1=m_2,m_2=m_3)(Pr(m_1=m_2)-Pr(m_2\neq m_3\mid m_1=m_2)Pr(m_1=m_2))
\end{split}
\end{equation*}

According to Eqs.~\ref{eq:impr_approx12} and~\ref{eq:impr_approx13}, we have $I_{m_1\rightarrow m_2}=Pr(m_1\neq m_2)$ and $I_{m_1\rightarrow m_3}-I_{m_1\rightarrow m_2} = Pr(m_2\neq m_3\mid m_1=m_2)Pr(m_1=m_2)$, so by substitution in the above equation we get:
\begin{equation*}
    Pr(m_1\neq m^{*},m_1=m_2,m_2=m_3)=Pr(m_1\neq m^{*}\mid m_1=m_2,m_2=m_3)(1 - I_{m_1\rightarrow m_3})
\end{equation*}

Similarly, the remaining terms in the Eq.~\ref{eq:impr_compute14} can be re-written as follows.
\begin{equation*}
\begin{split}
    Pr(m_1\neq m^{*},m_1=m_2,m_2\neq m_3)&=Pr(m_1\neq m^{*}\mid m_1=m_2,m_2\neq m_3)Pr(m_2\neq m_3\mid m_1=m_2)Pr(m_1=m_2) \\
    Pr(m_1\neq m^{*},m_1\neq m_2,m_2=m_3)&=Pr(m_1\neq m^{*}\mid m_1\neq m_2,m_2=m_3)Pr(m_2=m_3\mid m_1\neq m_2)Pr(m_1\neq m_2)\\
    Pr(m_1\neq m^{*},m_1\neq m_2,m_2\neq m_3)&=Pr(m_1\neq m^{*}\mid m_1\neq m_2,m_2\neq m_3)Pr(m_2\neq m_3\mid m_1\neq m_2)Pr(m_1\neq m_2)
\end{split}
\end{equation*}

Based on hypothesis~\ref{hypo:model_capability} and inferred possible model outputs as shown in Figure~\ref{fig:hypothesis}, we have $Pr(m_1\neq m^{*}\mid m_1=m_2,m_2\neq m_3)=1$ (observation 2), $Pr(m_1\neq m^{*}\mid m_1\neq m_2,m_2=m_3)=1$ (observation 3), and $Pr(m_1\neq m^{*}\mid m_1\neq m_2,m_2\neq m_3)=0$ (observation 4), so the equations above can be transformed as follows.
\begin{equation*}
\begin{split}
    Pr(m_1\neq m^{*},m_1=m_2,m_2\neq m_3)&=Pr(m_2\neq m_3\mid m_1=m_2)Pr(m_1=m_2)=I_{m_1\rightarrow m_3}-I_{m_1\rightarrow m_2}\\
    Pr(m_1\neq m^{*},m_1\neq m_2,m_2=m_3)&=Pr(m_2=m_3\mid m_1\neq m_2)Pr(m_1\neq m_2)\\
    Pr(m_1\neq m^{*},m_1\neq m_2,m_2\neq m_3)&=0
\end{split}
\end{equation*}
By substituting the above analysis results into Eq.~\ref{eq:impr_compute14}, we get:
\begin{equation*}
    I_{m_1\rightarrow m^{*}}\approx Pr(m_1\neq m^{*}\mid m_1=m_2,m_2=m_3)(1-I_{m_1\rightarrow m_3})+I_{m_1\rightarrow m_3}-I_{m_1\rightarrow m_2}+Pr(m_2=m_3\mid m_1\neq m_2)Pr(m_1\neq m_2)
\end{equation*}
We thus obtain Eq.~\ref{eq:impr_approx14}, completing the proof.
\section{Semantic Cardinality Estimator}\label{appx:cardinality_estimator}

\begin{figure}[h]
    \centering
    \includegraphics[width=\textwidth]{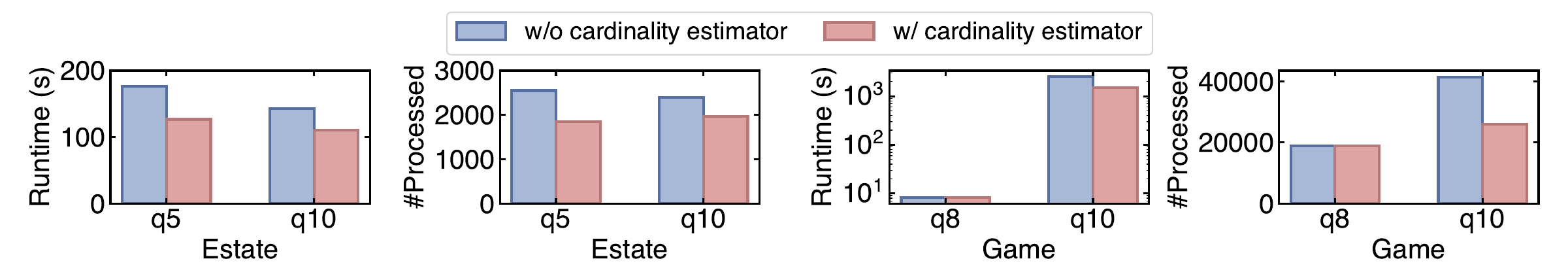}
    \vspace{-3mm}
    \caption{Impact of semantic cardinality estimation on the number of processed records and system latency.}
    \label{fig:cardinality_estimator}
    \vspace{-3mm}
\end{figure}

We evaluate the potential benefits of a semantic cardinality estimator. Since no such estimator currently exists, we simulate its effect using an oracle that provides the optimal execution order of operations in a query (by enumerating all possible query plans and selecting the best). We test this approach on \sql{q5} and \sql{q10} from Estate, and \sql{q8} and \sql{q10} from Game, where the optimal order of \sql{filter} operators is otherwise undetermined. Figure~\ref{fig:cardinality_estimator} shows that guiding the optimizer with a better operator order reduces the total number of records that \ours must process  by $20\%$ and $37\%$, thereby lowering system latency by $26\%$ and $40\%$ on Estate and Game, respectively.

\section{Batch Prompting}\label{appx:batch_prompting}

\begin{figure}[h]
    \centering
    \includegraphics[width=3.3in]{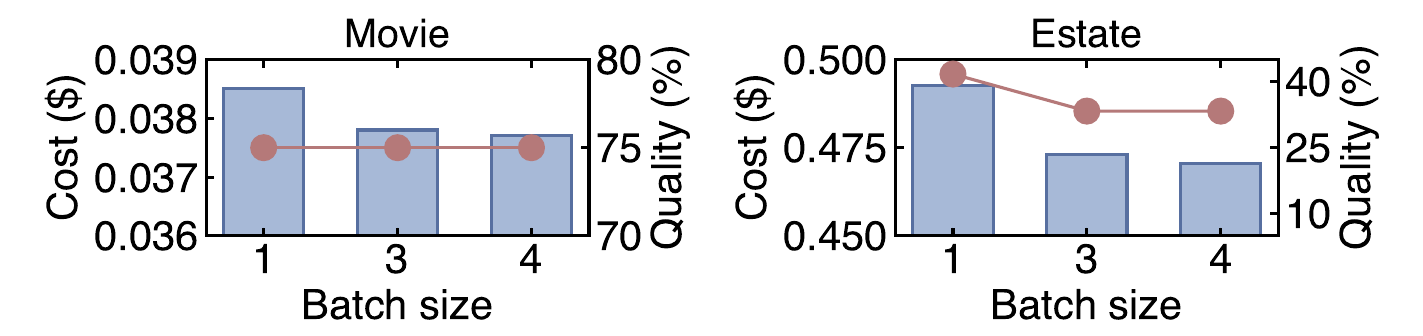}
    \vspace{-3mm}
    \caption{Impact of batch prompting on cost and quality.}
    \label{fig:batch_prompting}
    \vspace{-3mm}
\end{figure}

In this section, we examine the potential impact of batch prompting on semantic data analytics processing. Based on experiments from the original work~\cite{cheng2023batchprompting}, we compare \ours with no batch prompting (\ie batch$=1$) to \ours with batch prompting (batch$=3$ and batch$=4$, which provide a good trade-off between cost savings and accuracy according to the original paper). Figure~\ref{fig:batch_prompting} shows system performance in terms of cost savings and answer quality. As observed, the cost savings from batch prompting are modest: on Movie, batch prompting saves \$0.00069 and \$0.00078 for batch sizes 3 and 4, respectively; on Estate, savings are \$0.019 and \$0.022. However, the effect on processing quality varies across datasets: there is no degradation on Movie, while the quality drops from 42\% to 33\% on Estate. This is likely because Estate involves more complex analytical tasks than Movie.

\section{Additional Analysis on the Logical Optimizer}\label{appx:logical}

\begin{figure}[h]
\begin{minipage}[h]{0.5\linewidth}
    \centering
    \includegraphics[width=3.3in]{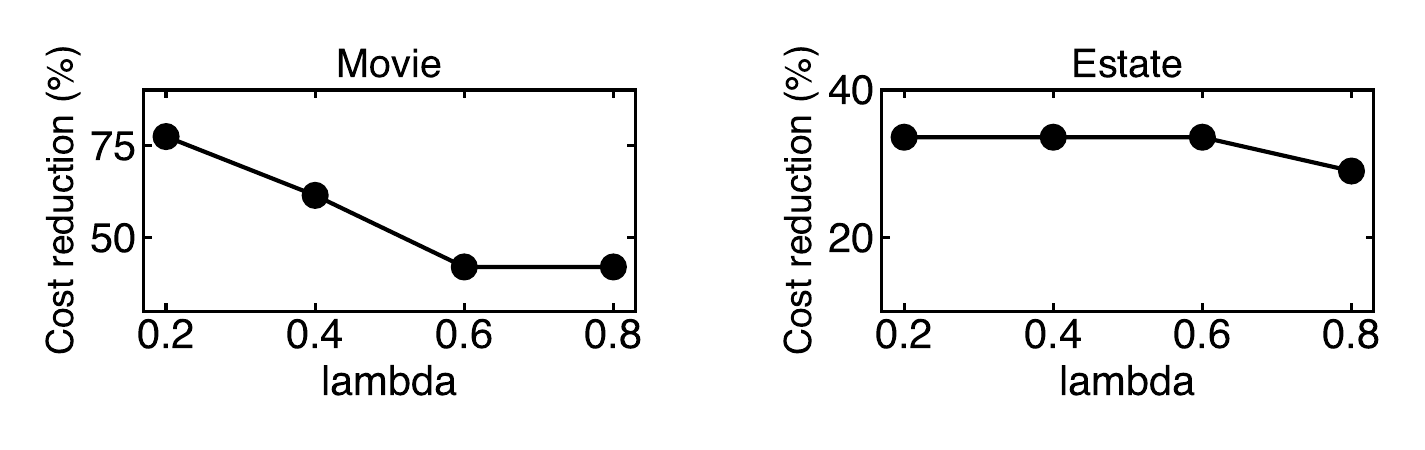}
    \vspace{-5mm}
    \caption{The sensitivity study on the parameter $\lambda$.}
    \label{fig:sensitivity}
\end{minipage}%
\begin{minipage}[h]{0.5\linewidth}
    \centering
    \setlength{\abovecaptionskip}{1.5mm}
    \captionof{table}{Comparison with beam search in terms of costs.}
    \renewcommand{\arraystretch}{1.1}
    \begin{tabular}{l c c c c}
    \toprule[1pt]
     & \multicolumn{2}{c}{\textbf{Movie}} & \multicolumn{2}{c}{\textbf{Estate}} \\
     & Optim. (\$) & Exec. (\$) & Optim. (\$) & Exec. (\$) \\ \hline
     \textsf{ours} & 0.0082 & 0.0304 & 0.0072 & 0.4017 \\
     \textsf{beam search} & 0.0133 & 0.0385 & 0.0122 & 0.4226 \\
    \bottomrule[1pt]
    \end{tabular}
    \label{tab:search_strategy}
\end{minipage}
\end{figure}

We first analyze the choice of $\lambda$. Figure~\ref{fig:sensitivity} presents a sensitivity study, where $\lambda$ is selected via a grid search over $[0,1]$ with a step size of 0.2. The parameter $\lambda$ has a varying impact on the logical optimizer across datasets. We choose $\lambda=0.2$ as it yields the largest cost reduction on most datasets.

We also compare our search strategy with a beam search that retains the two lowest-cost plans for optimization at each step. Table~\ref{tab:search_strategy} reports the results. Beam search incurs roughly double the optimization cost compared to our logical optimization approach, while the end-to-end system costs remain similar for both strategies.

\section{Risks and Challenges}\label{appx:challenge}
When applying \ours to some serious applications like medical data, there will be many risks and challenges that \ours needs to overcome. For instance,
\begin{itemize}[topsep=0pt,leftmargin=*]
    \item \textbf{Privacy issue.} When it comes to applications in the medical domains, the privacy of users would raise a great concern as the personal information and medical history would be leaked to the third-party models during interacting LLMs to processing analytical queries.
    \item \textbf{The use of context information.} For an accurate analysis, LLMs are required to process the context information such as relevant documents, long-/short-term memories. For now, our system are unable to process such context information. However, in the near future, we will support them via implementing RAG and multi-agents in our system.  
\end{itemize}

\section{Workloads}\label{appx:workloads}

\begin{lstlisting}[language=Python, caption=Workloads on the Movie dataset]
#Q1: Extract the genres of all movies.
df.semantic_map(user_instruction="According to the movie plot, extract the genre(s) of each movie.", input_column="Plot", output_column="Genre")

#Q2: Find all movies directed by Christopher Nolan.
df.semantic_filter(user_instruction="The movie is directed by Christopher Nolan.", input_column="Director")

#Q3: Find all movies whose poster is in the dark style.
df.semantic_filter(user_instruction="Whether the movie poster image is in the dark style.", input_column="Poster")

#Q4: Find all movie that has ever won more than 3 Oscars.
df.semantic_filter(user_instruction="Whether the movie has ever won more than 3 Oscars?", input_column="Awards")

#Q5: Compute the total box office gross of all movies whose ratings are higher than 9.
df.semantic_filter(user_instruction="The rating is higher than 9.", input_column="IMDB_rating")
df.semantic_reduce(user_instruction="Compute the total box office gross.", input_column="BoxOffice")

#Q6: Count the number of movies directed by Quentin Tarantino.
df.semantic_filter(user_instruction="The movie is directed by Quentin Tarantino.", input_column="Director")
df.semantic_reduce(user_instruction="Count the number of movies.", input_column="Title")

#Q7: Find the genre of the highest-rating movie directed by Steven Spielberg.
df.semantic_map(user_instruction="According to the movie plot, extract the genre(s) of each movie.", input_column="Plot", output_column="Genre")
df.semantic_filter(user_instruction="The movie is directed by Steven Spielberg.", input_column="Director")
df.semantic_reduce(user_instruction="Find the highest rate in the rest movie.", input_column="IMDB_rating")

#Q8: Count the number of movies that has won 2 Oscars with rating higher than 9.
df.semantic_filter(user_instruction="The rating is higher than 9.", input_column="IMDB_rating")
df.semantic_filter(user_instruction="Whether the movie has won 2 Oscars.", input_column="Awards")
df.semantic_reduce(user_instruction="Count the number of movies.", input_column="Title")

#Q9: Find the maximum rating of crime movie with a rating higher than 8.5 and lower than 9.
df.semantic_map(user_instruction="According to the movie plot, extract the genre(s) of each movie.", input_column="Plot", output_column="Genre")
df.semantic_filter(user_instruction="The rating is higher than 8.5.", input_column="IMDB_rating")
df.semantic_filter(user_instruction="The rating is lower than 9.", input_column="IMDB_rating")
df.semantic_filter(user_instruction="The movie belongs to crime movies.", input_column="Genre")
df.semantic_reduce(user_instruction="Find the maximum rating in the rest movies.", input_column="IMDB_rating")

#Q10: Count the number of crime movies with a rating higher than 8.5 and lower than 9.
df.semantic_map(user_instruction="According to the movie plot, extract the genre(s) of each movie.", input_column="Plot", output_column="Genre")
df.semantic_filter(user_instruction="The rating is higher than 8.5.", input_column="IMDB_rating")
df.semantic_filter(user_instruction="The rating is lower than 9.", input_column="IMDB_rating")
df.semantic_filter(user_instruction="The movie belongs to crime movies.", input_column="Genre")
df.semantic_reduce(user_instruction="Count the number of crime movies.", input_column="Title")

#Q11: Compute the average movie runtime of crime movies with a rating higher than 9.
df.semantic_map(user_instruction="According to the movie plot, extract the genre(s) of each movie.", input_column="Plot", output_column="Genre")
df.semantic_filter(user_instruction="The rating is higher than 9.", input_column="IMDB_rating")
df.semantic_filter(user_instruction="The movie belongs to crime movies.", input_column="Genre")
df.semantic_reduce(user_instruction="Compute the average movie runtime.", input_column="Runtime")

#Q12: Extract the main character of crime movies with rating higher than 9.
df.semantic_map(user_instruction="According to the movie plot, extract the genre(s) of each movie.", input_column="Plot", output_column="Genre")
df.semantic_filter(user_instruction="The rating is higher than 9.", input_column="IMDB_rating")
df.semantic_filter(user_instruction="The movie belongs to crime movies.", input_column="Genre")
df.semantic_map(user_instruction="Extract the main character from the movie plot.", input_column="Plot", output_column="Character")
\end{lstlisting}

\begin{lstlisting}[language=Python, caption=Workloads on the Estate dataset]
#Q1: Find the house with a yard.
df.semantic_filter(user_instruction="Observed from the house picture, whether the house has a yard or not.", input_column="image")

#Q2: Extract the house price from details.
df.semantic_map(user_instruction="Extract the house price from the detail about the estate.", input_column="Details", output_column="Price")

#Q3: Whether the house is located in Ajah, Lagos.
df.semantic_filter(user_instruction="Whether the house is located in Ajah, Lagos.", input_column="Location")

#Q4: Extract amenities of the estate.
df.semantic_map(user_instruction="Extract Amenities of the estate from the estate details.", input_column="Details", output_column="Amenities")

#Q5: Extract amenities of the estate that has more than 3 bedrooms but less than 6 bedrooms.
df.semantic_filter(user_instruction="Whether the estate has more than 3 bedrooms", input_column="Title")
df.semantic_map(user_instruction="Extract Amenities of the estate from the estate details.", input_column="Details", output_column="Amenities")
df.semantic_filter(user_instruction="Whether the estate has less than 6 bedrooms.", input_column="Title")

#Q6: Compute the average price for the estates that seem to have a yard.
df.semantic_map(user_instruction="Extract the house price from the detail about the estate.", input_column="Details", output_column="Price")
df.semantic_filter(user_instruction="Observed from the house picture, whether the house has a yard or not.", input_column="image")
df.semantic_reduce(user_instruction="Compute the average price for the estates.", input_column="Price")

#Q7: Extract features of the estates that have 2 or 3 bedrooms from the estate details.
df.semantic_map(user_instruction="Extract features from the detail about the estate.", input_column="Details", output_column="Features")
df.semantic_filter(user_instruction="Whether the estate has 2 or 3 bedrooms", input_column="Title")

#Q8:  Extract amenities of the estates that have 2 or 3 bedrooms from the estate details.
df.semantic_map(user_instruction="Extract amenities from the detail about the estate.", input_column="Details", output_column="Amenities")
df.semantic_filter(user_instruction="Whether the estate has 2 or 3 bedrooms", input_column="Title")

#Q9: Compute the average price for the estates that has more than 3 bedrooms but less than 6 bedrooms.
df.semantic_map(user_instruction="Extract the house price from the detail about the estate.", input_column="Details", output_column="Price")
df.semantic_filter(user_instruction="Whether the estate has more than 3 bedrooms", input_column="Title")
df.semantic_filter(user_instruction="Whether the estate has less than 6 bedrooms.", input_column="Title")
df.semantic_reduce(user_instruction="Compute the average price for the estates.", input_column="Price")

#Q10: Compute the lowest price for the detached duplex that has more than 3 bedrooms and less than 6 bedrooms.
df.semantic_map(user_instruction="Extract the house price from the detail about the estate.", input_column="Details", output_column="Price")
df.semantic_filter(user_instruction="Whether the estate has more than 3 bedrooms.", input_column="Title")
df.semantic_filter(user_instruction="Whether the estate has less than 6 bedrooms.", input_column="Title")
df.semantic_filter(user_instruction="Whether the estate is a detached duplex.", input_column="Title")
df.semantic_reduce(user_instruction="Compute the lowest price for the estates.", input_column="Price")

#Q11: Compute the lowest price for the estates that has a swimming pool.
df.semantic_map(user_instruction="Extract the house price from the detail about the estate.", input_column="Details", output_column="Price")
df.semantic_map(user_instruction="Extract the amenities from the estate details.", input_column="Details", output_column="Amenities")
df.semantic_filter(user_instruction="Is there a swimming pool in the estate.", input_column="Amenities")
df.semantic_reduce(user_instruction="Compute the lowest price for the estates.", input_column="Price")

#Q12: Compute the average price for the estates with a gym and a swimming pool and located in Lekki, Lagos.
df.semantic_map(user_instruction="Extract the house price from the detail about the estate.", input_column="Details", output_column="Price")
df.semantic_map(user_instruction="Extract the amenities from the estate details.", input_column="Details", output_column="Amenities")
df.semantic_filter(user_instruction="Is there a swimming pool in the estate.", input_column="Amenities")
df.semantic_filter(user_instruction="Is there a gym in the estate.", input_column="Amenities")
df.semantic_filter(user_instruction="Is the estate located in Lekki, Lagos.", input_column="Location")
df.semantic_reduce(user_instruction="Compute the average price for the estates.", input_column="Price")
\end{lstlisting}

\begin{lstlisting}[language=Python, caption=Workloads on the Steam dataset]
#Q1: Select the games for adults.
df.semantic_filter(user_instruction="According to the given PEGI rating (in picture), check if the game is only suitable for adults (18 years or older).", input_column="rating")

#Q2: Convert overall review to a binary label.
df.semantic_map(user_instruction="Give the video game a binary review (positive or negative) based on the existing review.", input_column="overall_reviews", output_column="comments")

#Q3: Does the video game support VR.
df.semantic_filter(user_instruction="Does the video game support VR?", input_column="platforms")

#Q4: Find all games with a MetaCritic score higher than 90.
df.semantic_filter(user_instruction="The rating is higher than 90.", input_column="metacriticts")

#Q5: Find the publisher that produces sports video games.
df.semantic_map(user_instruction="Extract the genre from the brief summary of the game.", input_column="description", output_column="genre")
df.semantic_filter(user_instruction="The video game is about sports.", input_column="genre")
df.semantic_reduce(user_instruction="Find the publisher that appears the most.", input_column="publisher")

#Q6: Compute the lowest discounted price of the video game that support MacOS.
df.semantic_filter(user_instruction="Is MacOS in the list of supported platforms?", input_column="platforms")
df.semantic_reduce(user_instruction="Find the lowest price.", input_column="discounted_price")

#Q7: Find all shooting games that support Chinese language.
df.semantic_map(user_instruction="Extract the genre from the brief summary of the game.", input_column="description", output_column="genre")
df.semantic_filter(user_instruction="The video game is about shooting.", input_column="genre")
df.semantic_filter(user_instruction="Is Chinese one of the supported languages?", input_column="language")

#Q8: Count the number of games that only have one developer with a rating higher than 90.
df.semantic_filter(user_instruction="The rating is higher than 90.", input_column="metacriticts")
df.semantic_filter(user_instruction="Does the video game has only one developer?", input_column="developer")
df.semantic_reduce(user_instruction="Count the number of games.", input_column="title")

#Q9: Compute the average price in USD of games that support VR and are shooting games.
df.semantic_map(user_instruction="Extract the genre from the brief summary of the game.", input_column="description", output_column="genre")
df.semantic_filter(user_instruction="Does the game support VR.", input_column="platforms")
df.semantic_filter(user_instruction="The game is a shooting game", input_column="genre")
df.semantic_map(user_instruction="Convert the price in IDR into the price in USD.", input_column="discounted_price", output_column="price_usd")
df.semantic_reduce(user_instruction="Compute the average price in USD of games.", input_column="price_usd")

#Q10: Compute the average price of games that support both Windows and MacOS and receive a positive review.
df.semantic_map(user_instruction="Give the video game a binary review (positive or negative) based on the existing review.", input_column="overall_reviews", output_column="comments")
df.semantic_filter(user_instruction="The game receives a positive review.", input_column="comments")
df.semantic_filter(user_instruction="According to the given PEGI rating (in picture), check if the game is only suitable for adults (18 years or older).", input_column="rating")
df.semantic_filter(user_instruction="The game supports both Windows and MacOS.", input_column="platforms")
df.semantic_reduce(user_instruction="Compute the average original price.", input_column="original_price")

#Q11: Find the earliest release date of a game that supports VR.
df.semantic_filter(user_instruction="The game supports VR.", input_column="platforms")
df.semantic_reduce(user_instruction="Find the earliest release date.", input_column="release_date")

#Q12: Find the earliest release date of an adventure game with a metacritic score higher than 90.
df.semantic_filter(user_instruction="The rating is higher than 90.", input_column="metacriticts")
df.semantic_map(user_instruction="According to the cover image of the video game, summarize its graphic style.", input_column="image", output_column="graphic_style")
df.semantic_filter(user_instruction="The graphic style of the game is cartoon.", input_column="graphic_style")
df.semantic_reduce(user_instruction="Find the earliest release date.", input_column="release_date")
\end{lstlisting}

\end{document}